\documentclass[twocolumn,amssymb, nobibnotes, aps, pra, superscriptaddress]{revtex4-2}

\usepackage{hyperref}
\usepackage{amsmath}
\usepackage{braket}
\usepackage{bm}
\usepackage{graphicx}
\usepackage{float} 
\usepackage{subfigure} 
\usepackage{color}
\usepackage{ulem}
\usepackage[ruled,linesnumbered]{algorithm2e}

\begin{document}

\title{Quantum Gaussian filter for exploring ground state properties}
\author{Min-Quan He}
\affiliation{Guangdong-Hong Kong Joint Laboratory of Quantum Matter, Department of Physics, and HKU-UCAS Joint Institute for Theoretical and Computational Physics at Hong Kong, The University of Hong Kong, Pokfulam Road, Hong Kong, China}
\author{Dan-Bo Zhang}
\email{dbzhang@m.scnu.edu.cn}
\affiliation{Guangdong-Hong Kong Joint Laboratory of Quantum Matter, Frontier Research Institute for Physics,
South China Normal University, Guangzhou 510006, China}
\affiliation{Guangdong Provincial Key Laboratory of Quantum Engineering and Quantum Materials, School of Physics
and Telecommunication Engineering, South China Normal University, Guangzhou 510006, China}

\author{Z. D. Wang}
\email{zwang@hku.hk}
\affiliation {Guangdong-Hong Kong Joint Laboratory of Quantum Matter, Department of Physics, and HKU-UCAS Joint Institute for Theoretical and Computational Physics at Hong Kong, The University of Hong Kong, Pokfulam Road, Hong Kong, China}

\begin{abstract}
Filter methods realize a projection from a superposed quantum state onto a target state, which can be efficient if two states have sufficient overlap. Here we propose a quantum Gaussian filter (QGF) with the filter operator being a Gaussian function of system Hamiltonian. A hybrid quantum-classical algorithm feasible on near-term quantum computers is developed, which implements the quantum Gaussian filter as a linear combination of Hamiltonian evolution at various times. Remarkably, the linear combination coefficients are determined classically and can be optimized in the post-processing procedure. Compared to the existing filter algorithms whose coefficients are given in advance, our method is more flexible in practice under given quantum resources with the help of post-processing on classical computers. We demonstrate the quantum Gaussian filter algorithm for the quantum Ising model with numeral simulations under noises. We also propose an alternative full quantum approach that implements QGF with an ancillary continuous-variable mode.
\end{abstract}

\maketitle

\section{Introduction}
\label{Introduction}
Solving ground-state problems plays a significant role in quantum simulation~\cite{georgescu2014quantum, seth1996universal}, which can be applied to many fields such as quantum chemistry and condensed-matter physics~\cite{mcardle2020quantum, aspuru2005simulated, lanyon2011universal, greiner2002quantum}. The ground-state property is estimated as an expectation value of the ground state for an observable operator~\cite{zhang2021computing, meir1992landauer, jensen2017introduction, o2019calculating}. As classical algorithms would eventually encounter the exponential wall issue when solving larger-size quantum systems due to the exponentially increasing matrix dimension, quantum computing offers a way forward that dramatically decreases the resource required. Quantum phase estimation~(QPE), a milestone quantum algorithm for eigensolvers, uses a quantum Fourier transformation to extract eigenvalues and associated eigenstates probabilistically from an initial state~\cite{kitaev1995quantum, abrams1997simulation, aspuru2005simulated, dorner2009optimal, Abrams1999quantum}. However, QPE relies on long coherent evolution and can be resource-consuming for the era of noisy intermediate-scale quantum~(NISQ) devices~\cite{preskill2018quantum}. Unlike QPE, variational quantum eigensolvers~(VQEs) target the desired eigenstate with a parametrized quantum circuit~\cite{kandala2017hardware, liu2019variational, nakanishi2019subspace, wang2019accelerated, zhang2020collective,Amaro_2022}, whose parameters can be optimized with a hybrid quantum-classical procedure. While the circuit depth can be much reduced, VQE strongly depends on the ansatz~\cite{tang2021qubit}, and the optimizing process may be hindered by vanishing gradients known as barren plateaus~\cite{mcclean2018barren, cerezo2021cost}.

In principle, rethinking quantum phase estimation treats all eigenstates as equal, and the distribution of results is determined by the overlapping between each eigenstate and the initial state. However, once focusing on eigenstates of physical interest, such as the ground state and low-lying excited states, a projection onto such an eigenstate may be much more efficient. The key point is to construct a filter operator on a quantum computer to distinguish the desired eigenstate from the initial state. As the unwanted components of eigenstates are filtered out and their information can be ignored, quantum filter methods do not require quantum Fourier transformation; thus, it may be realized with many fewer quantum resources.

Recently, there have been several proposals for quantum filters, including the cosine-filtering operator~\cite{ge2019faster, lu2021algorithms}, the quantum inverse iteration algorithm~\cite{kyriienko2020quantum,he2022inverse}, the quantum filter diagonalization~\cite{parrish2019quantum}, and the powered Hamiltonian approximation~\cite{bespalova2021hamiltonian, seki2021quantum}. Those methods construct the nonunitary filter operator as a linear combination of unitaries implemented by a series of controlled unitaries or with a classical weighted summation. The coefficients are given beforehand~\cite{ge2019faster,kyriienko2020quantum,bespalova2021hamiltonian, seki2021quantum}, enjoying a clear analysis of desired resources, but they may not be adjustable on noisy quantum devices or need to be obtained by optimization~\cite{parrish2019quantum} in a high-dimensional space. A compromise of two may make a quantum filter both theoretically promised and flexible for implementation on near-term quantum devices. It is required to design a proper quantum filter with some hyperparameters that can be fittingly optimized in classical postprocessing.

In this work we propose a quantum Gaussian filter (QGF) for exploring the ground-state properties of quantum systems. The QGF solves the approximate ground state by performing a Gaussian function of the Hamiltonian on a given initial state which has a sufficient overlapping with the ground state. The Gaussian filter is parameterized with a shift energy and a width, which can be adjusted for optimally distinguishing singling out the ground state. We present a hybrid quantum-classical algorithm for implementing the QGF, optimizing the Gaussian filter operator in the classical postprocessing. Compared to the existing algorithms whose filter operators are fixed by the preset coefficients~\cite{ge2019faster,kyriienko2020quantum,bespalova2021hamiltonian, seki2021quantum}, our algorithm can construct an optimal filter operator by classical optimization under limited quantum resources. This feature makes this QGF algorithm suitable on NISQ devices. We demonstrate the algorithm by numerically simulating the quantum Ising model under noises. In addition, we give a full quantum version of the QGF with an auxiliary continuous variable using only Gaussian states. This continuous-variable-assisted approach provides an alternative way to implement the Gaussian filter eigensolver on hybrid qubit-quantum mode (qumode) quantum computers. It only requires the entanglement between qubits and a single ancillary qumode instead of plenty of ancillary qubits while constructing the linear combination of unitaries.

This paper is organized as follows. We first introduce our algorithm's principle and procedure in Sec.~\ref{Sec_Quantum_algorithm} and analyze its time complexity in Sec.~\ref{Sec_Time_complexity_analysis}. Then the numerical results of this algorithm are shown in Sec.~\ref{Sec_Demonstration}. A continuous-variable-assisted strategy is presented in Sec.~\ref{Sec_Continuous_variable_assisted_algorithm}. A summary and discussion are given in Sec.~\ref{Sec_Conclusion_and_discussion}.

\begin{figure*}
    \centering
        \begin{minipage}[b]{0.9\textwidth}
            \includegraphics[width=1\textwidth]{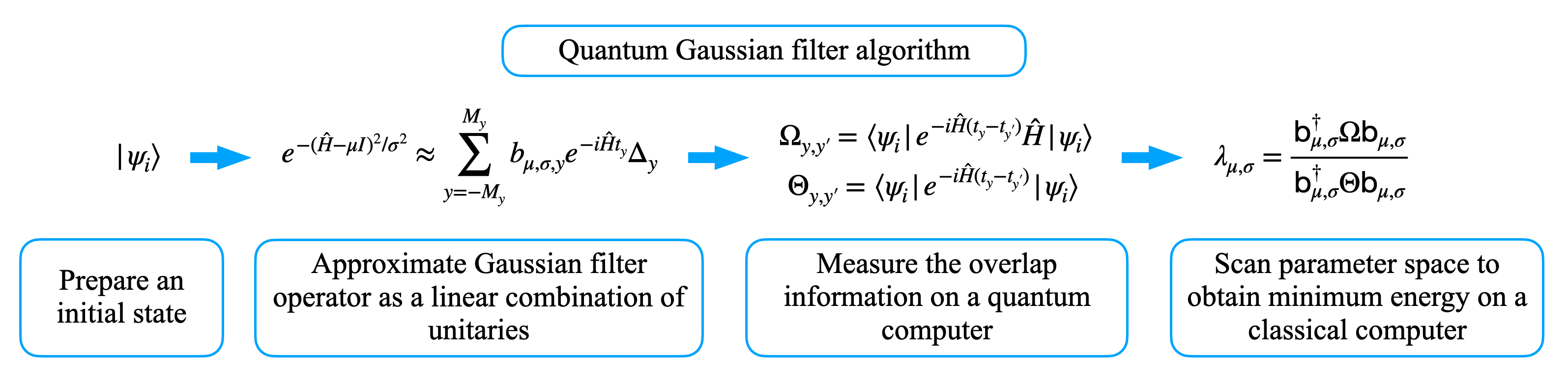}
        \end{minipage} 
    \caption{The QGF algorithm starts from preparing an initial quantum state and approximates the Gaussian filter operator as a linear combination of unitary operators. The next step is to evolve the initial state and measure the overlap matrices on a quantum computer. The estimated energy is a classical summation of the overlap information with corresponding weight. Finally, the minimum result is obtained by scanning the parameter space on a classical computer.}  \label{procedure}
\end{figure*}

\section{Quantum algorithm}
\label{Sec_Quantum_algorithm}
This section presents the principle and procedures of the QGF algorithm. We consider a gapped local Hamiltonian $\hat{H} = \sum_{l = 1}^{L} c_{l} \hat{h}_{l}$, where $\hat{h}_{l}$ is a Pauli string performing on a subset of qubits and $L$ is a polynomial of the system size $N$. The local Hamiltonian can well describe the system in many disciplines, including quantum chemistry~\cite{mcardle2020quantum, aspuru2005simulated}, quantum many-body simulation~\cite{georgescu2014quantum}, and combinatorial optimization problems~\cite{farhi2014quantum}. The main idea of this algorithm is to perform a Gaussian filter operator $e^{-(\hat{H} - \mu I)^{2}/\sigma^{2}}$ on an arbitrary initial state $\ket{\psi_{i}}$ that has nonzero overlap with the exact ground state, where the expected value $\mu$ and variance $\sigma^{2}$ of the Gaussian function correspond to a negative shift energy and the width of the Gaussian filter. By representing the initial state as a superposition of the eigenstate of an $N$-qubit Hamiltonian $\hat{H}$, e.g. $\ket{\psi_{i}} = \sum_{j = 0}^{2^{N} - 1} a_{j} \ket{\lambda_{j}}$, a Gaussian filter operator $e^{-(\hat{H} - \mu \hat{I})^{2}/\sigma^{2}}$ converts it to
\begin{equation}
    \ket{\psi_{f}} = \frac{1}{\sqrt{C}} \sum_{j = 0}^{2^{N} - 1} a_{j} e^{- (\lambda_{j} - \mu)^{2}/\sigma^{2}} \ket{\lambda_{j}},
\end{equation}
where $\frac{1}{\sqrt{C}} = (\sum_{j = 0}^{2^{N} - 1} a_{j}^{2} e^{- (\lambda_{j} - \mu)^{2}/\sigma^{2}})^{-1/2}$ is the normalization factor, $\ket{\lambda_{j}}$ is the eigenstate of $\hat{H}$ corresponding to $j$th smallest eigenvalue $\lambda_{j}$, and $a_{j} = |\braket{\lambda_{j}|\psi_{i}}|$ is the square root of the overlap between the initial state and eigenstate $\ket{\lambda_{j}}$. This Gaussian filter operator $e^{-(\hat{H} - \mu I)^{2}/\sigma^{2}}$ generates an additional weight $e^{-(\lambda_{j} - \mu)^{2}/\sigma^{2}}$ for each eigenstate, which monotonically decreases with the eigenvalue. If all the eigenvalues are positive values, as the variance $\sigma^{2}$ or the expected value $\mu$ of the Gaussian function decreases, the resulting state converges to the ground state $\ket{\lambda_{0}}$.

Since the Gaussian filter operator $e^{-(\hat{H} - \mu \hat{I})^{2}/\sigma^{2}}$ is a nonunitary operator that cannot naturally be performed on quantum computers, it is approximated by a linear combination of unitary operators~\cite{Berry2015, low2017optimal, berry2015simulating, Low2019hamiltonian} with the idea of a Fourier approximation
\begin{equation} \label{eq:appro_gaussian_filter}
    e^{-(\hat{H} - \mu \hat{I})^{2}/\sigma^{2}}  \approx  \sum_{y = -M_{y}}^{M_{y}} b_{\mu, \sigma, y} e^{- i \hat{H} t_{y}} \Delta_{y}\equiv G_{\mu,\sigma}(\hat{H}),
\end{equation}
where $\Delta_{y}$ is the slice size, $M_{y}$ is the cutoff number of terms of unitary operators, $t_{y} = y \Delta_{y}$ is the evolution time of the $y$th term, 
\[b_{\mu, \sigma, y} = \frac{\sigma}{2 \sqrt{\pi}} e^{- (y \Delta_{y} \sigma)^{2}/4} e^{ i \mu (y \Delta_{y})} \]
is the weight of the $y$th term, and the maximum evolution time is defined as $\phi_{m} = M_{y} \Delta_{y}$. 
Performing the above operator $G_{\mu,\sigma}(\hat{H})$ on an arbitrary initial state leads to an additional weight $g_{\sigma, \mu}(\lambda)$ for every eigenstate $\ket{\lambda}$ that is
\begin{equation}
    \label{Gaussion_function_of_eigenvalue}
    g_{\mu, \sigma}(\lambda_{j}) = \sum_{y = -M_{y}}^{M_{y}} b_{\mu, \sigma, y} e^{- i \lambda_{j} t_{y}} \Delta_{y}.
\end{equation}
The above equation can well approximate $e^{- (\lambda - \mu)^{2}/\sigma^{2}}$ for $\lambda \in [\mu, \lambda_{m} + \mu]$ and oscillates around $e^{- (\lambda_{m})^{2}/\sigma^{2}}$ for $\lambda > \lambda_{m} + \mu $ for $\phi_{m} = 2 \lambda_{m}/\sigma^{2}$ and $\Delta_{y} = O(1/\lambda_{m})$ (see Appendix~\ref{Sec_Approximation_error_of_Gaussian_filter} for details).

The approximate Gaussian filter can be constructed by the linear combination of unitaries (LCU) lemma~\cite{ge2019faster}, which encodes the coefficients into the ancillary qubits and applies the Hamiltonian evolution controlled by the ancillary qubits. However, it may require a deep quantum circuit. Instead, we directly solve the ground-state energy by sequential energy estimation~\cite{kyriienko2020quantum}. It helps to avoid a large number of controlled Hamiltonian evolution, which can significantly decrease the requirement of a quantum resource. Another reason is that it can separate coefficients into the classical parts to scan the parameters of the Gaussian filter classically. As mentioned above, a perfect Gaussian filter with lower $\sigma$ and $\mu$ results in higher accuracy since the excited states are filtered more. However, it is harder to implement and more sensitive to quantum noise. In this case, a classical optimization and quantum error mitigation strategy can help determine the best-performance coefficient under limited quantum resources.

The approximate ground-state energy is estimated as $\lambda_{\mu, \sigma} = \braket{\psi_{\mu, \sigma}|\hat{H}|\psi_{\mu, \sigma}}/\braket{\psi_{\mu, \sigma}|\psi_{\mu, \sigma}}$.
Considering the sequential energy estimation method~\cite{kyriienko2020quantum} that classically sums the overlap between the initial and the evolved state with corresponding weight, the estimated ground-state energy is
\begin{equation}
    \label{Eq:Classical_summation_of_expectation_value}
    \begin{aligned}
        \lambda_{\mu, \sigma}
        =&  \frac{\braket{\psi_{\mu, \sigma} | \hat{H} | \psi_{\mu, \sigma}}}{\braket{\psi_{\mu, \sigma} | \psi_{\mu, \sigma}}} \\
        =& \frac{\sum_{y, y^{\prime} = -M_{y}}^{M_{y}} b_{\mu, \sigma, y} b_{\mu, \sigma, y^{\prime}}^{\ast} \braket{\psi_{i}|\hat{H} e^{-i (t_{y} - t_{y^{\prime}}) \hat{H}}|\psi_{i}}}{\sum_{y, y^{\prime} = -M_{y}}^{M_{y}} b_{\mu, \sigma, y} b_{\mu, \sigma, y^{\prime}}^{\ast} \braket{\psi_{i}|e^{-i (t_{y} - t_{y^{\prime}}) \hat{H}}|\psi_{i}}}.
    \end{aligned}
\end{equation}
As the parameters of the Gaussian filter operator are determined by the choice of classical weights $\{ b_{\mu, \sigma, y} \}_{y = -M_{y}}^{M_{y}}$, the estimated energy can be optimized by classical post-processing without external quantum computation.

As the Gaussian filters are nonunitary operators, the resulting state after a Gaussian filter is applied should be renormalized, i.e., $\ket{\psi_{\mu, \sigma}} = G_{\mu, \sigma} \ket{\psi_{i}}/\lVert G_{\mu, \sigma} \ket{\psi_{i}}\rVert$. It requires a post-selection process for the LCU lemma, where $\lVert G_{\mu, \sigma} \ket{\psi_{i}}\rVert$ is the successful probability of preparing the resulting state. The successful probability may decay while lowering $\sigma$ and $\mu$. However, the problem of decay does not exist in our method. Instead, the denominator of Eq.~\eqref{Eq:Classical_summation_of_expectation_value} plays the role of renormalization. A new issue arises that the scales of both the denominator and numerator decay as $\sigma$ and $\mu$ decrease. Under this situation, the result is more sensitive to statistical error and quantum noise. Therefore, it is essential to set an appropriate range of $\sigma$ and $\mu$ to scan depending on the number of query samples and the noise (details are shown in Sec.~\ref{Sec_Time_complexity_analysis}).

The procedure of the QGF algorithm to solve an approximate ground state energy is shown in Fig.~\ref{procedure}, whose steps are as follows.
\begin{itemize}
    \item[1.] Set the approximation parameters $M_{y}$ and $\Delta_{y}$ according to the system size of the Hamiltonian to be solved and the desired accuracy.
    \item[2.] Prepare an initial state $\ket{\psi_i}$ with a nonzero overlap with the exact ground state. 
    \item[3.] Evaluate the overlap information in Eq.~\eqref{Eq:Classical_summation_of_expectation_value} $\braket{\psi_{i}|\hat{H} e^{-i (t_{y} - t_{y}^{\prime}) \hat{H}}|\psi_{i}}$ and $\braket{\psi_{i}|e^{-i (t_{y} - t_{y}^{\prime}) \hat{H}}|\psi_{i}}$ for $y, y^{\prime} \in [-M_{y},M_{y}]$ on a quantum processor.
    \item[4.] Optimize the classical weights $\{ b_{\mu, \sigma, y} \}_{y = -M_{y}}^{M_{y}}$ by  minimizing the approximate ground state energy $\lambda_{\mu, \sigma}$ on a classical computer.
\end{itemize}

In step 3 the overlap measurement can be completed by the Hadamard test~\cite{aharonov2009polynomial} with an ancillary qubit. The denominator term is an overlap between the initial state and the evolved state. It starts from the ancillary qubit and register qubits being initialized as $\frac{1}{\sqrt{2}} (\ket{0} + \ket{1})$ and $\ket{\psi_{i}}$. After a time evolution $e^{-i (t_{y} - t_{y}^{\prime}) \hat{H}}$ is applied to the register qubits controlled by the ancillary qubit and a Hadamard gate applied to the ancillary qubit, the expected value of the result by measuring the ancillary qubit is Re$(\braket{\psi_{i}|e^{-i (t_{y} - t_{y}^{\prime}) \hat{H}}|\psi_{i}})$. Similarly, its imaginary part can be obtained by initializing the ancillary qubit as $\frac{1}{\sqrt{2}} (\ket{0} - i\ket{1})$. There exists a Hamiltonian for the numerator term. It can be represented in the following form by decomposing the Hamiltonian into local operators:
\begin{equation}
    \braket{\psi_{i}|\hat{H} e^{-i (t_{y} - t_{y}^{\prime}) \hat{H}}|\psi_{i}} = \sum_{l = 1}^{L} c_{l} \braket{\psi_{i}|\hat{h}_{l} e^{-i (t_{y} - t_{y}^{\prime}) \hat{H}}|\psi_{i}}.
\end{equation}
Each term in this above equation can be measured in the same way as the denominator term but applied to an additional controlled Pauli string $\hat{h}_{l}$ after the controlled Hamiltonian evolution, and a classical weighted summation computes the numerator term. In this overlap measurement process, the time evolution of the Hamiltonian is decomposed into local Pauli rotation by Trotter decomposition~\cite{suzuki1991general} to be implemented on quantum computers,
\begin{equation}
    e^{-i (t_{y} - t_{y}^{\prime}) \hat{H}} \approx \left[\prod_{l = 1}^{L} e^{-i c_{l} \hat{h}_{t} (t_{y} - t_{y}^{\prime})/n}\right]^{n}.
\end{equation}

The energy estimation procedure can also be used as an iterative strategy. It starts from certain level approximation parameters $M_{y}$ and $\Delta_{y}$ to obtain an estimated ground state energy. We may iteratively change the approximation parameters to increase the result accuracy until it meets the requirement. In each iteration, only additional overlaps in Eq.~\eqref{Eq:Classical_summation_of_expectation_value} are required to be measured. Moreover, this algorithm can also estimate the ground state properties $\braket{\psi_{g}|\hat{A}|\psi_{g}}$ with an observable operator $\hat{A}$.
Once the best-performance parameters $\{ b_{\mu, \sigma, y} \}_{y = -M_{y}}^{M_{y}}$ are obtained, any expectation value of the ground-state property $\braket{\psi_{g}|\hat{A}|\psi_{g}}$ can be solved by replacing $\hat{H}$ with the observable operator $\hat{A}$ in Eq.~\eqref{Eq:Classical_summation_of_expectation_value}.

\section{Time complexity analysis}
\label{Sec_Time_complexity_analysis}

We briefly discuss the time complexity of the QGF algorithm for estimating the ground-state energy of a given gapped local Hamiltonian within a fixed accuracy. The time complexity contains the circuit depth and query complexity. Recalling the local Hamiltonian assumption $\hat{H} = \sum_{l = 1}^{L} c_{l} \hat{h}_{l}$, the number of local operators $L$ is a polynomial of the system size. This indicates that the range of the energy spectrum $\lambda_{2^{N}-1} - \lambda_{0}$ is also a polynomial of the system size. Since we do not normalize the Hamiltonian to constrain the energy spectrum, the spectral gap is $O(1)$ for a gapped local Hamiltonian.

Let us first consider the maximum circuit depth, which is determined by the maximum evolution time. The initial state can be written as a superposition of the target ground state and its orthogonal state $\ket{\psi_{i}} = a_{0} \ket{\lambda_{0}} + \sqrt{1-a_{0}^{2}} \ket{\lambda_{\perp}}$, where $a_{0}$ is the absolute value of the square root of the overlap between the initial state and the ground state and $\ket{\lambda_{\perp}}$ is the state perpendicular to the ground state. We construct a QGF by a linear combination of unitaries with a maximum evolution time $\phi_{m} = 2\lambda_{m}/\sigma^{2}$. Applying this QGF on the eigenstate $\ket{\lambda_{j}}$ leads to the additional weight $g_{\mu, \sigma} (\lambda_{j})$ shown in Eq.~\ref{Gaussion_function_of_eigenvalue}, which can well approximate a Gaussian function $e^{-(\lambda_{j} - \mu)^{2}/\sigma^{2}}$ for $\lambda_{j} \in [\mu, \mu + \lambda_{m}]$ and oscillates around $e^{-(\lambda_{m} - \mu)^{2}/\sigma^{2}}$ for $\lambda_{j} > \lambda_{m}$. Setting the shift energy $\mu = \lambda_{0} + \Delta - \lambda_{m}$, the QGF applied to the ground state and the other states lead to additional weights $e^{-(\lambda_{m} - \Delta)^{2}/\sigma^{2}}$ and $e^{-\lambda_{m}^{2}/\sigma^{2}}$, respectively. The resulting state is 
\[\ket{\psi_{f}} = \frac{1}{\sqrt{C}} [a_{0} e^{-(\lambda_{m} - \Delta)^{2}/\sigma^{2}} \ket{\lambda_{0}} + \sqrt{1-a_{0}^{2}} e^{-\lambda_{m}^{2}/\sigma^{2}} \ket{\lambda_{\perp}}],\] where 
\[1/\sqrt{C} = [a_{0}^{2} e^{-2(\lambda_{m} - \Delta)^{2}/\sigma^{2}} + (1 - a_{0}^{2}) e^{-2\lambda_{m}^{2}/\sigma^{2}}]^{-1/2}\]
is the normalization factor.
The estimated energy then is $E_{e} = E_{g} + O(a_{0}^{-2} e^{(2 \Delta^{2} - 4 \Delta \lambda_{m})/\sigma^{2}})$. As the energy gap is independent of the system size, the requirement of $\lambda_{m}$ for a fixed accuracy $\epsilon$ is $O(\sigma^{2}\log(a_{0}^{-2} \epsilon^{-1}))$, corresponding to a maximum evolution time $\phi_{m} =  2\lambda_{m}/\sigma^{2} = O(\log(a_{0}^{-2} \epsilon^{-1}))$. It is a logarithm of inverse accuracy and overlap between the initial state and the ground state. After analyzing the maximum evolution time, we discuss its corresponding circuit depth.

The evolution of the Hamiltonian $e^{-i \hat{H} t}$ is implemented by Trotterization~\cite{suzuki1991general}, i.e., $e^{-i \hat{H} t} \approx (\prod_{l = 1}^{L} e^{-i c_{l} \hat{h}_{t} t/n})^{n}$.
The circuit depth of Trotter decomposition is $O(L^{3} c_{max} t^{2}/\epsilon)$ for time evolution $e^{-i\hat{H}t}$ and the state accuracy is within $\epsilon$. Here, we neglect the maximum weight of the local operator $c_{max}$ since it is $O(1)$. Since the combination effects of the scale of the denominator of Eq.~\eqref{Eq:Classical_summation_of_expectation_value} are $O(a_{0}^{-2} e^{-2\lambda_{m}^{2}/\sigma^{2}})$ and the amplitude of the classical weight for the $e^{-i\hat{H}t}$ term is $O(\sigma e^{-t^{2}\sigma^{2}/4})$, the gate complexity of $e^{-i \hat{H} t}$ is $O(a_{0}^{-2} \epsilon^{-1} L^{3} t^{2} \sigma e^{2\lambda_{m}^{2}/\sigma^{2}} e^{-t^{2} \sigma^{2}/4})$, where the absolute value of the evolution time ranges from $0$ to $2\phi_{m} = 4\lambda_{m}/\sigma^{2}$. Therefore, if $\sigma \leq 2\lambda_{m}$, the maximum value of the circuit complexity is $O(a_{0}^{-2} \epsilon^{-1} L^{3} \sigma^{-1} e^{2\lambda_{m}^{2}/\sigma^{2}})$ for $t = 2/\sigma$; if $\sigma > 2\lambda_{m}$, the maximum value of the circuit complexity is $O(a_{0}^{-2} \epsilon^{-1} L^{3} \lambda_{m}^{2} \sigma^{-3} e^{\lambda_{m}/\sigma^{2}})$ for $t = 4\lambda_{m}/\sigma^{2}$ (details are shown in Appendix~\ref{Sec_Requirement_of_Trotter_number_and_statistical_accuracy}). We set an appropriate scale for $\sigma = \frac{1}{2} \log^{-1} (a_{0}^{-2} \epsilon^{-1})$, the required value of $\lambda_{m}$ being $\sigma^{2} (a_{0}^{-2} \epsilon^{-1}) = \frac{1}{4} \log^{-1} (a_{0}^{-2} \epsilon^{-1})$. Under this setting, the circuit depth is $O(a_{0}^{-2} \epsilon^{-1} L^{3} \log(a_{0}^{-2} \epsilon^{-1}))$, which is a polynomial of inverse initial overlap, desired accuracy, and system size.

We continue to discuss the query complexity of this algorithm, which includes the number of terms and query times required for each term in Eq.~\eqref{Eq:Classical_summation_of_expectation_value}. To approximate $e^{-(\lambda - \mu)^{2}/\sigma^{2}}$ for $\lambda \in [\mu, \mu + \lambda_{m}]$, the slice fineness of the Fourier approximation $\Delta_{y}$ should be $O(1/\lambda_{m})$. Similarly to the error analysis of Trotter decomposition, the requirement of statistical error for each term is also determined by the scale of denominator and coefficient amplitude. There are $4M_{y} + 1$ unique items in the denominator, where the range of evolution time is $t \in [-2\phi_{m}, 2\phi_{m}]$, and the interval is $\Delta y$. For the $y$-th term, the number of query samples should be $O(\epsilon^{-1} \sigma a_{0}^{-2} e^{2\lambda_{m}^{2}/\sigma^{2}} e^{-(y\Delta y)^{2} \sigma^{2}/4})$. Therefore, the query complexity is $O(\epsilon^{-1} a_{0}^{-2} \lambda_{m} e^{2\lambda_{m}^{2}/\sigma^{2}})$ for solving a result with accuracy $\epsilon$ (Details are shown in Appendix~\ref{Sec_Requirement_of_Trotter_number_and_statistical_accuracy}). For the same set of $\sigma$ above, the query complexity is $O(\epsilon^{-1} a_{0}^{-2} \log^{-1}(a_{0}^{-2} \epsilon^{-1}))$. It is a polynomial of inverse initial overlap, desired accuracy, and system size.

In total, the time complexity of this QGF algorithm is $O(\epsilon^{-2} L^{3} a_{0}^{-4})$ for solving an approximate ground-state energy with a desired accuracy $\epsilon$. The time complexity is a polynomial of system size, the inverse of error, and the inverse of initial overlap. 
It should be noted that our result depends on the initial state. A random initial state may cause the overlap to exponentially decay as the system size increases, leading to exponential time complexity. This phenomenon widely exists in the quantum filter algorithms~\cite{ge2019faster,kyriienko2020quantum, he2022inverse, bespalova2021hamiltonian, seki2021quantum} and QPE algorithms~\cite{kitaev1995quantum, aspuru2005simulated, dorner2009optimal, Abrams1999quantum, alan2005simulated}. 
It meets the theoretical expectation since the local Hamiltonian problem is a QMA-hard problem.
Preparing an appropriate initial state may help improve the performance of this kind of algorithm to eliminate this issue slightly. It may be a possible way to prepare the initial state by exploiting  the prior knowledge of the Hamiltonian.
Another potential way is to prepare the initial state by a VQE with a shallow depth quantum circuit. Only when the initial overlap is a polynomial of system size will the QGF algorithm have polynomial time complexity.

\section{Demonstration}
\label{Sec_Demonstration}
In this section, we simulate the QGF algorithm to solve the approximate ground state energy of a transverse-field Ising model, whose Hamiltonian is
\begin{equation}
    \label{eq:Hamiltonian}
    \hat{H} = -J \sum_{n = 1}^{N} \hat{\sigma}_{n}^{z} \hat{\sigma}_{n + 1}^{z} + g \sum_{n = 1}^{N} \hat{\sigma}_{n}^{x},
\end{equation}
where $J$ is the interaction strength of nearby sites, $g$ is the scale of the external transverse field, and $\hat{\sigma}^{z}_{N+1} = \hat{\sigma}^{z}_{1}$ for periodic boundary conditions. 
The simulation is conducted  with the package QUTIP~\cite{JOHANSSON20131234}. The open-source code for this demonstration is available on GitHub~\cite{he2022github}. In the numerical demonstration, the overlap evaluation is implemented with a quantum circuit.
The properties of the QGF are investigated by exploring different parameters. Remarkably, we simulate the QGF under several quantum noise models to show that the QGF can be a potential candidate as a NISQ quantum algorithm.

\begin{figure}
    \centering
    \subfigure[]{
        \begin{minipage}[b]{0.45\textwidth}
            \includegraphics[width=1\textwidth]{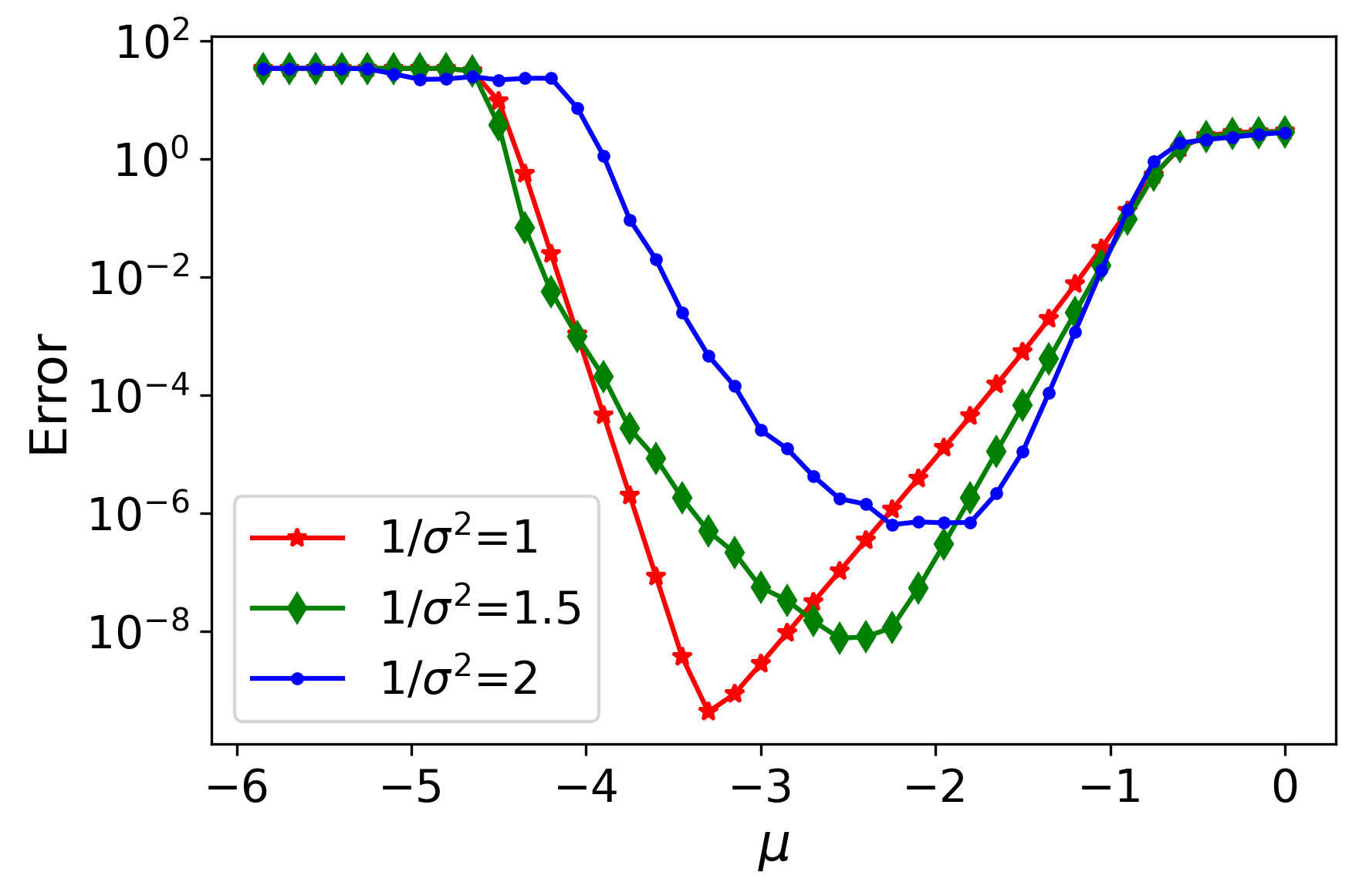}
        \end{minipage}
    }
    \subfigure[]{
        \begin{minipage}[b]{0.45\textwidth}
            \includegraphics[width=1\textwidth]{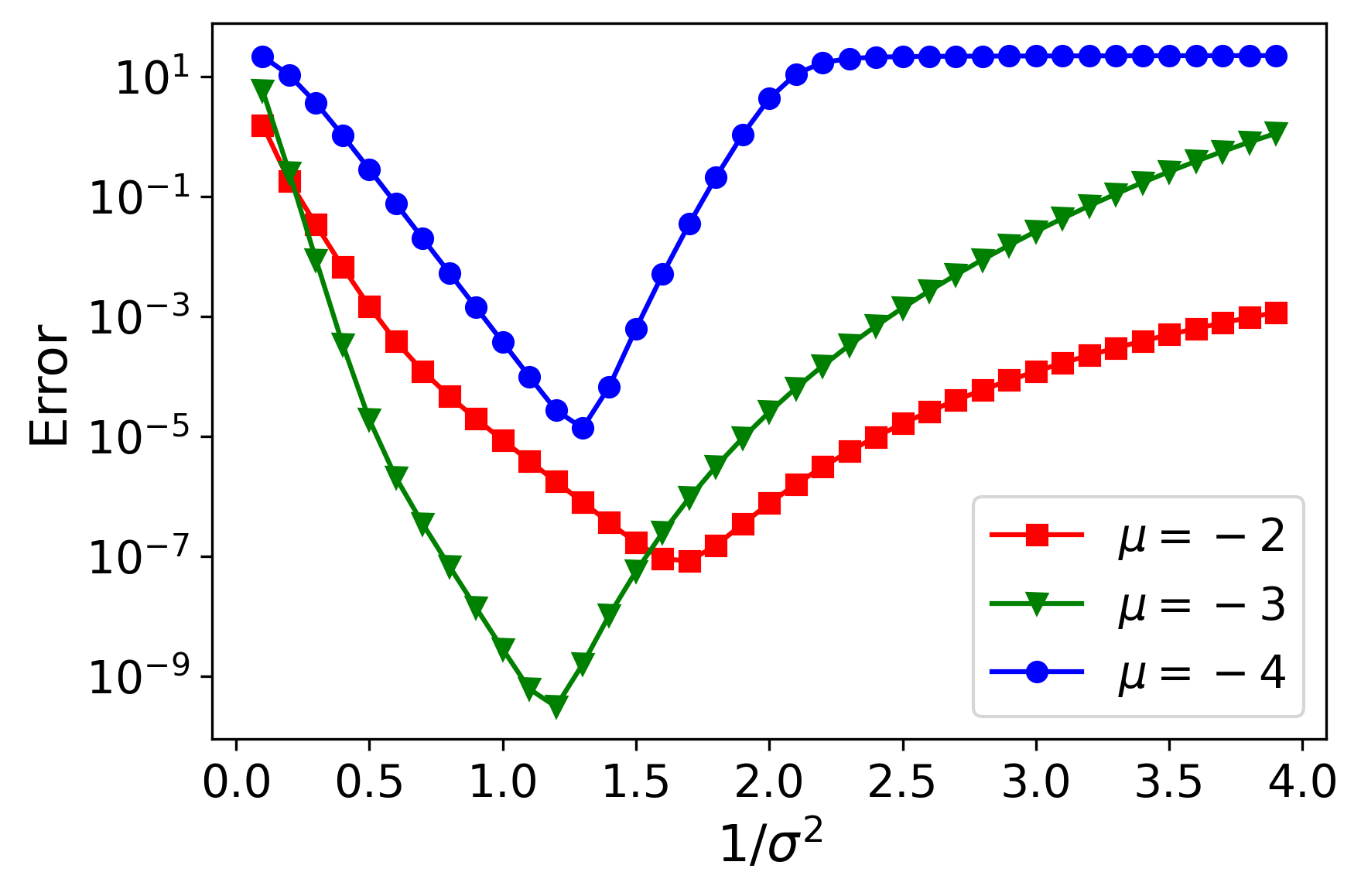}
        \end{minipage}
    }
    \caption{Demonstration of the ground-state energy estimation by the QGF algorithm for a transverse-field Ising model with $N = 8$ qubits. The discrete parameters are set as $\Delta_{y} = 0.16$ and $M_{y} = 50$. (a) Energy between the exact ground state energy and the estimated energy for different negative shift energy $\mu$. Red, green, and blue lines correspond to the cases $1/\sigma^{2} = 1$, $1.5$, and $2$, respectively. (b) Energy difference as a function of $1/\sigma^{2}$. Red, green, and blue lines correspond to the cases $\mu = -2$, $-3$, and $-4$, respectively.}
    \label{Ising_phase_8}
\end{figure}

We first consider solving the transverse-field Ising model with $J = 1$, $g = 2$, and $N = 8$ for fixed discrete parameters $\Delta_{y} = 0.16$ and $M_{y} = 50$. A shift energy of $15$ is applied to shift all the eigenvalues, and a random initial state is used. As shown in Fig.~\ref{Ising_phase_8}, the result error first decreases and then increases as $\mu$ ($\sigma^{2}$) decreases for a fixed $\sigma^{2}$ ($\mu$). This results from the Gaussian filter with a smaller width and negative shift energy having a better performance but being more difficult to approximate. The result also shows that even starting from a random initial state with a low overlap with the ground state, a high accuracy result can be reached by this QGF algorithm with a maximum evolution time $\phi_{m} = 8$.

\begin{figure}
    \centering

    \begin{minipage}[b]{0.45\textwidth}
        \includegraphics[width=1\textwidth]{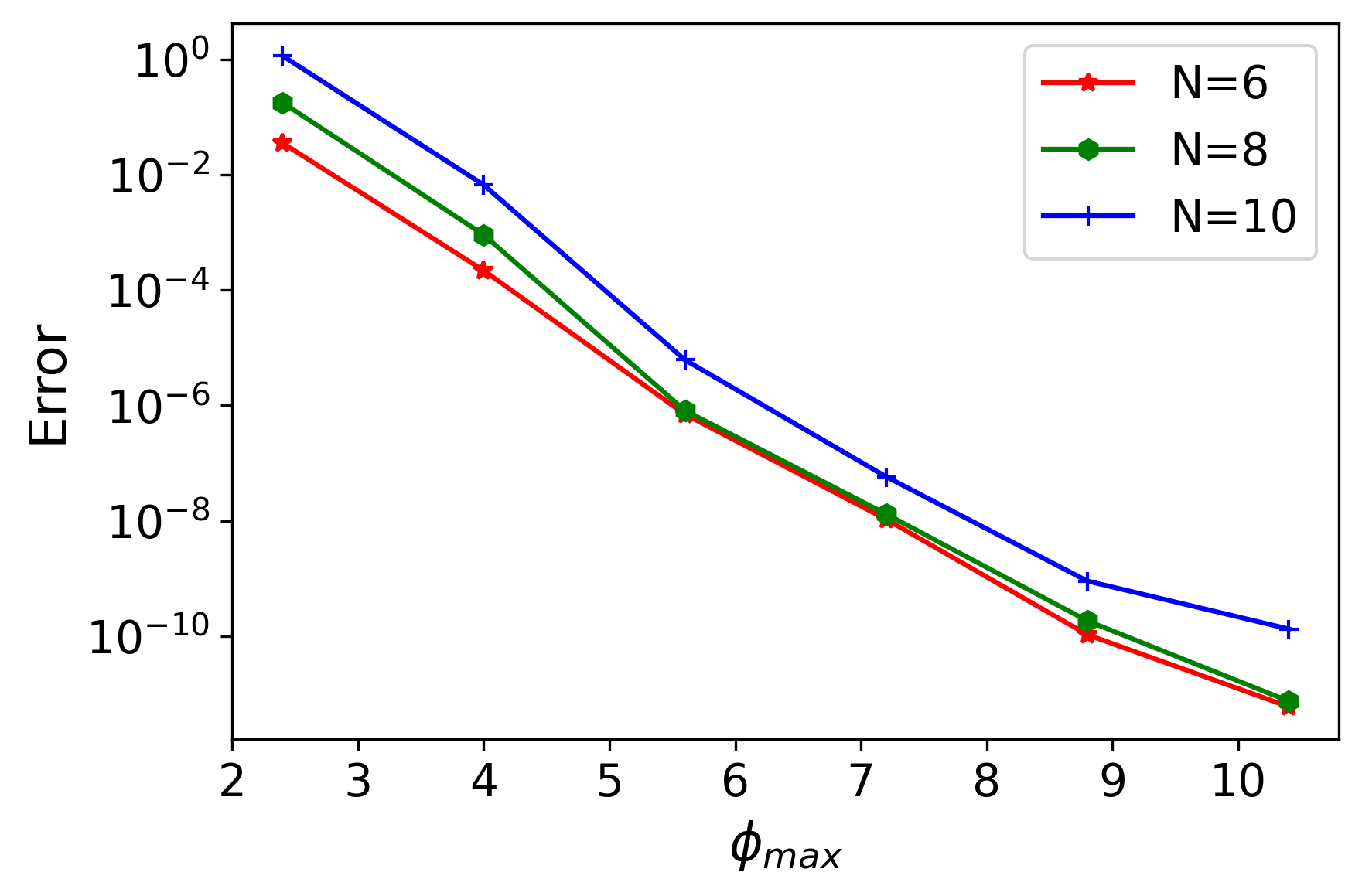}
    \end{minipage}
  
    \caption{Energy difference between estimated ground-state energy and the exact ground state energy as a function of maximum evolution time. The red, green, and blue lines correspond to the cases of $N = 6$, $8$, and $10$,  respectively.}
      \label{Ising_N_6_8_10_E_vs_phase}
\end{figure}

We also demonstrate the solution of the above problem for $N = 6$, $8$, and $10$ by an iterative approach, where $\Delta_{y} = 0.08$ and $M_{y}$ gradually increases from $30$ to $130$, corresponding to $\phi_{m} \in [2.4, 10.4]$. New overlap information is computed in each iteration, and minimum energy is obtained by searching for $1/\sigma^{2} \in [0.1, 3]$ and $\mu \in [\lambda_{0}, \lambda_{0} - 1]$. The result is shown in Fig.~\ref{Ising_N_6_8_10_E_vs_phase}, where the energy difference between the exact ground state and the approximate one solved by this iterative method decreases as the maximum evolution time increases for all three models.

\begin{figure}
    \centering
    \subfigure[]{
    \begin{minipage}[b]{0.45\textwidth}
        \includegraphics[width=1\textwidth]{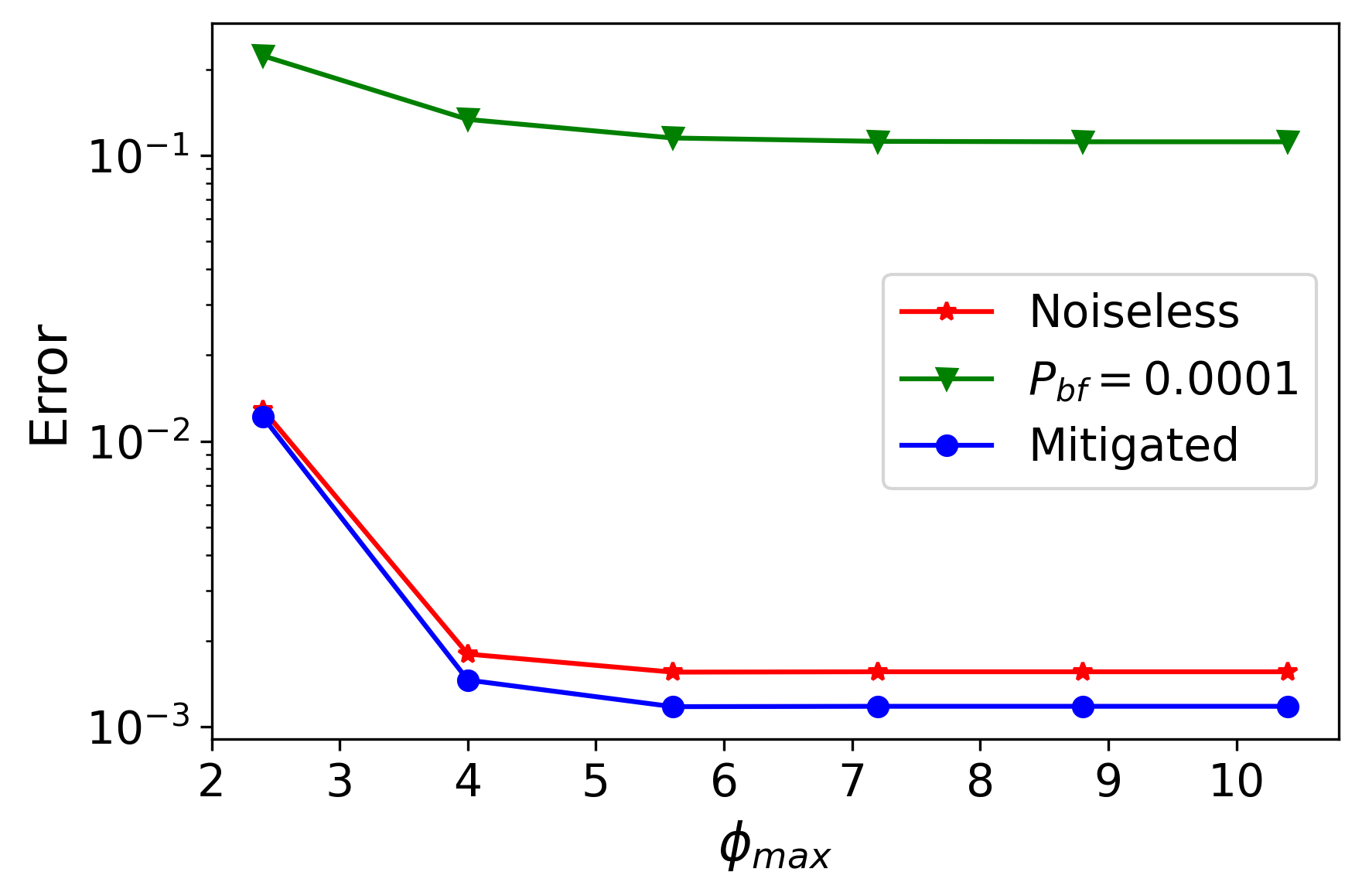}
        \label{Ising_N_4_Noise_E_vs_phase_bf}
    \end{minipage}}
    \subfigure[]{
    \begin{minipage}[b]{0.45\textwidth}
        \includegraphics[width=1\textwidth]{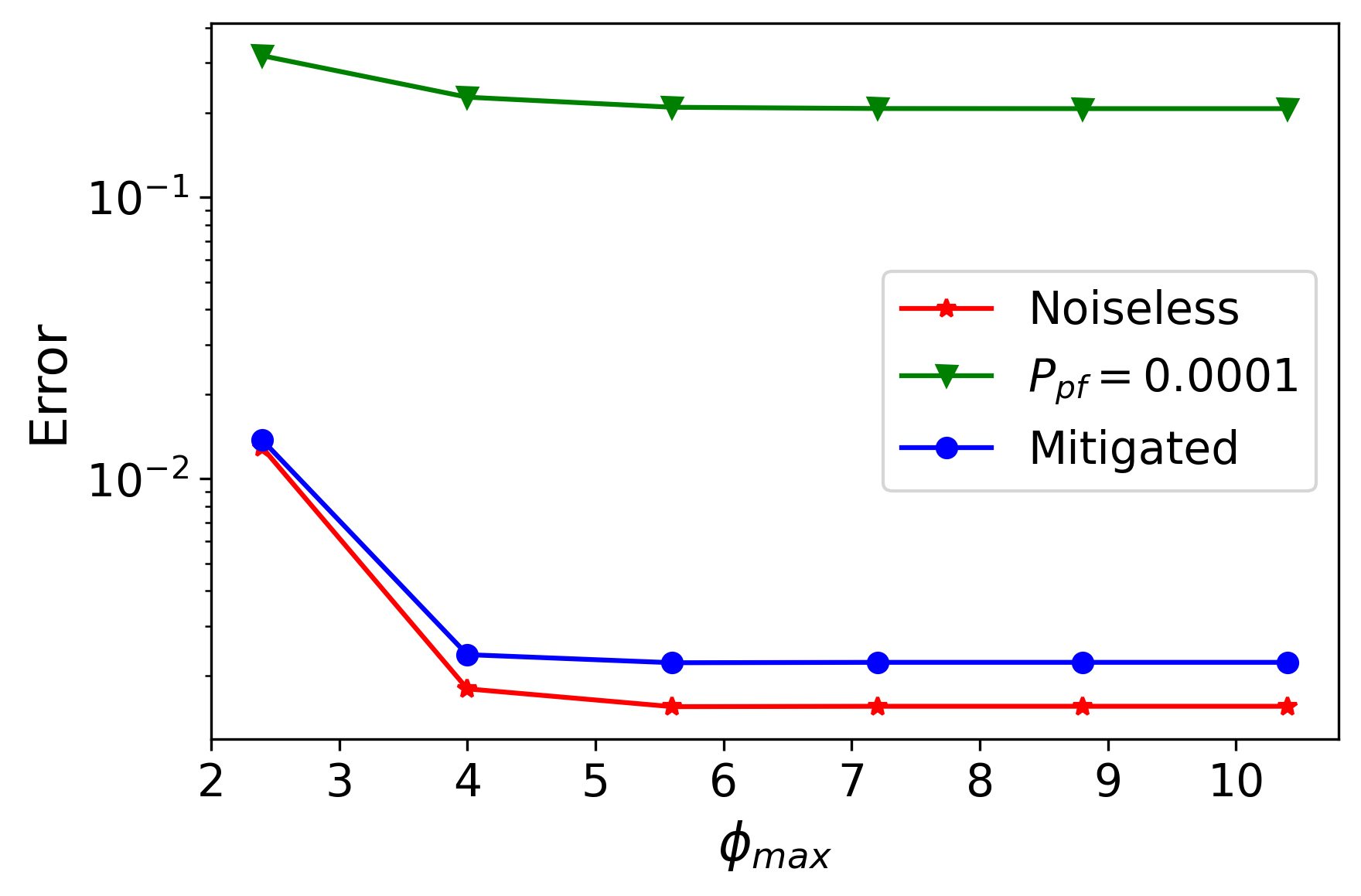}
        \label{Ising_N_4_Noise_E_vs_phase_pf}
    \end{minipage}}
    
    \caption{Numerical results for solving the approximate ground-state energy of the transverse-field Ising model with $N = 4$ at different maximum evolution times in a noisy environment. For both figures, red and green lines correspond to the noiseless and noisy cases, respectively; the blue line is the result under noise, but a noise mitigation strategy is used. (a) Solution of the approximate ground-state energy using a QGF for bit-flip noise. (b) Solution of the approximate ground-state energy using a QGF for phase-flip noise.}
    \label{Ising_N_4_Noise_E_vs_phase}
\end{figure}

While universal fault-tolerant quantum computers have not been developed yet, noises on near-term quantum processors should be considered. We demonstrate the quantum Gaussian filter algorithm by solving the above problem with $N = 4$ under noises. Here, Trotter decomposition~\cite{suzuki1991general} is applied for decomposing the Hamiltonian evolution into a sequence of one-qubit and two-qubit gates. We consider two noise models, the bit-flip and the phase-flip channels, which are applied for all qubits after each gate with a probability of $p = 0.0001$. The Trotter step is chosen as $20$ for single discrete-time evolution $\Delta_{y} = 0.08$. As the long evolution term suffers a more significant noise effect but has a lower weight, this quantum algorithm may naturally resist some noise influence. As shown in Fig.~\ref{Ising_N_4_Noise_E_vs_phase}, the result is affected by the noise since the noise scale is large. However, this can still be mitigated to a lower level by error mitigation using the zero-limit extrapolation method~\cite{temme2017error}, which is well suitable for short-depth quantum circuits. 

We also compare the QGF with other filter-based quantum algorithms. One candidate is the cosine filter algorithm, which is well developed and considered powerful for solving the ground state problem~\cite{lu2021algorithms,ge2019faster}. As a demonstration, we present a comparison between the Gaussian filter and the cosine filter in Appendix~\ref{Sec_Comparison_to_the_cosine_filter_algorithm}. We compare the approximation effect of the Gaussian function of eigenvalues. Remarkably, we apply the postprocess optimization into the cosine filtering operator to solve a ground-state energy. The results show that they have a similar performance. Furthermore, the results also show that the proposed classical optimization process can be adapted to other filtering operators.

\begin{figure}
    \centering
    \subfigure[]{
    \begin{minipage}[b]{0.45\textwidth}
        \includegraphics[width=1\textwidth]{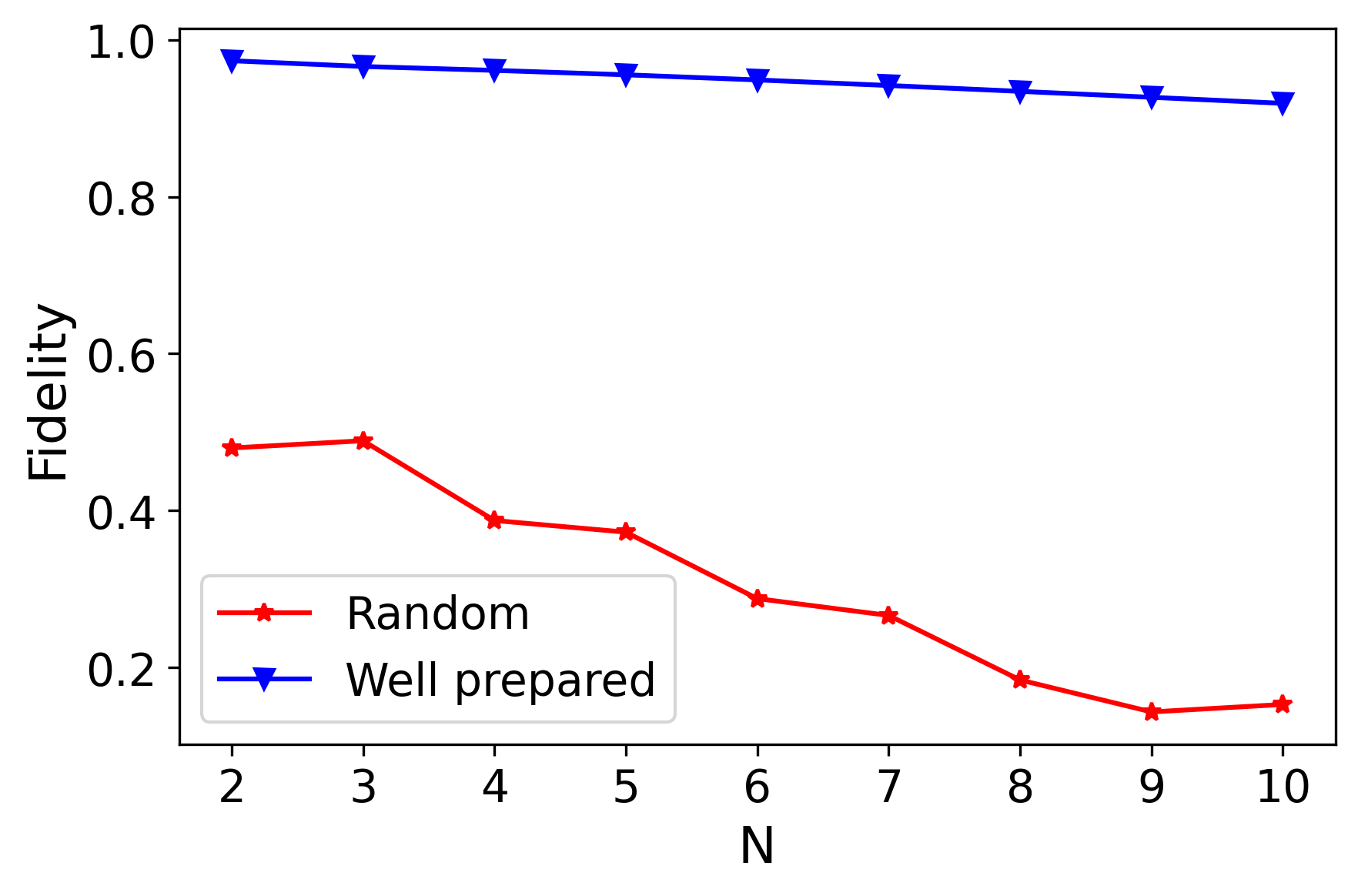}
        \label{ini_state_Z_domian}
    \end{minipage}}
    \subfigure[]{
    \begin{minipage}[b]{0.45\textwidth}
        \includegraphics[width=1\textwidth]{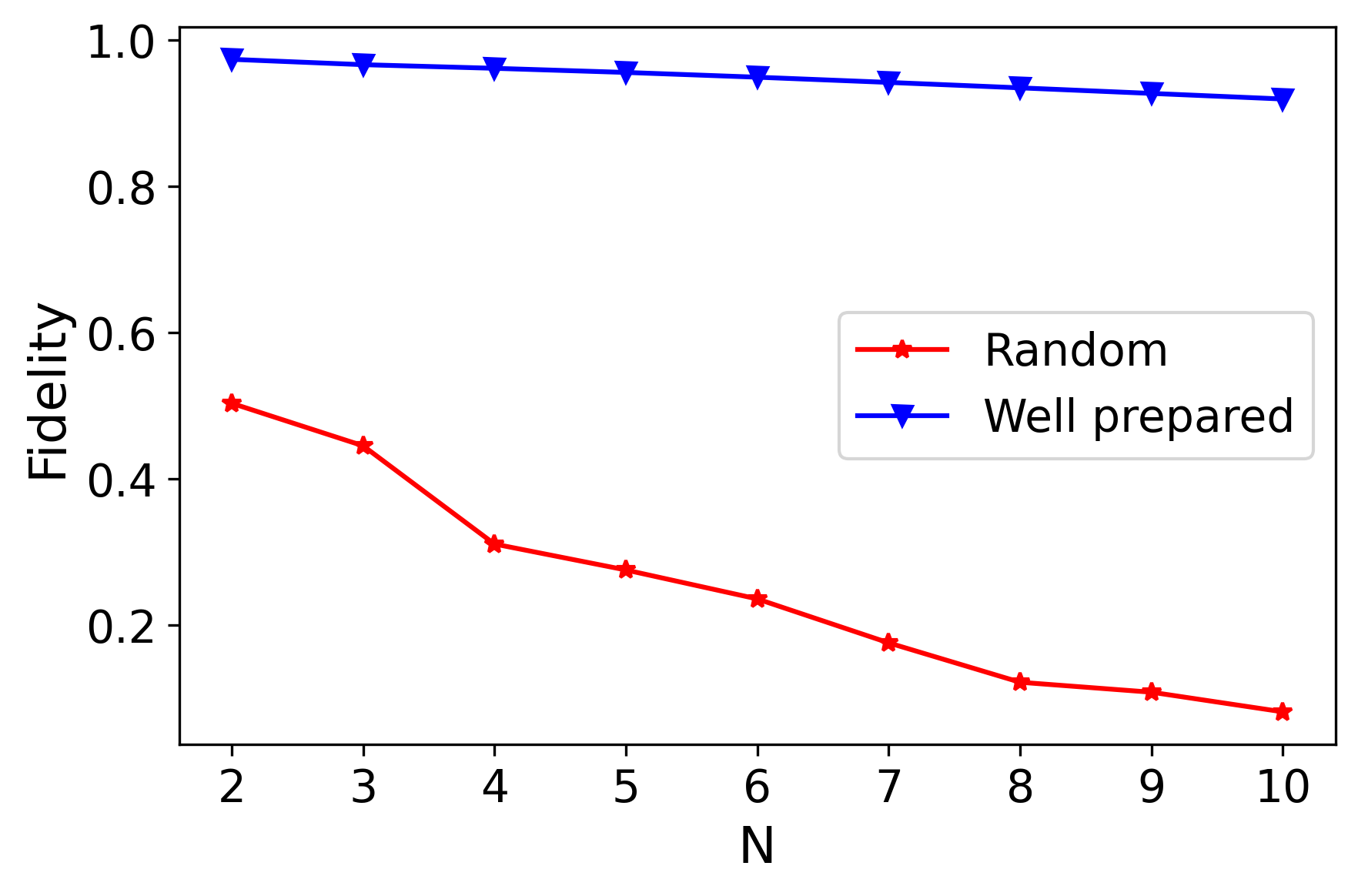}
        \label{ini_state_X_domian}
    \end{minipage}}
    
    \caption{Fidelity between the initial state and the ground state as a function of the number of qubits $N$. The red line represents the average result of $50$ randomly generated initial states. The blue line corresponds to the initial state well prepared with prior knowledge of the system Hamiltonian. (a) Preparation of the initial state for solving the transverse-field Ising model with $J/g=2$. (b)Preparation of the initial state for solving the transverse-field Ising model with $g/J=2$.}
    \label{ini_state}
\end{figure}

As analyzed in Sec.~\ref{Sec_Time_complexity_analysis}, the performance of the QGF algorithm depends on the initial state preparation. Instead of adopting a random initial state, preparing an initial state with prior knowledge of the system Hamiltonian may increase the performance of this quantum algorithm in practice. Here we give an example of preparing a high-quality initial state of the transverse-field Ising model with the help of physical intuition. If the transverse field is relatively small $|J| \gg |g|$, the first term in Eq.\eqref{eq:Hamiltonian} dominates. As a result, the corresponding ground state will be either all spin up or all spin down with a high probability. If a strong transverse field is applied, $|g| \gg |J|$, the ground state has a large overlap with the ground state of the second term. Therefore, we can prepare the initial state by the following procedure. When $|J| > |g|$, we first prepare a Greenberger-Horne-Zeilinger state by applying the Hadamard gate and controlled-NOT gate and then applying a phase-flip gate on each qubit; when $|J| < |g|$, we initialize each qubit as a ground state of $\hat{\sigma}^{x}$ by applying a Pauli-$Z$ gate after a Hadamard gate on $\ket{0}$. We numerically show the fidelity of the prepared state and the exact ground state as a function of system size, where the following two cases are considered: $J/g = 2$ and $g/J = 2$. As a comparison, we also prepare $50$ random initial states by rotating the state $\ket{0}^{\otimes N}$ through $N$ layers of a quantum alternating operator ansatz~\cite{hadfield2019quantum} by random angles and show the average fidelity. The random initial state is represented as $\ket{\psi_{i}(\vec{\beta}, \vec{\gamma})} = [\prod_{j=1}^{N} e^{-i \beta_{j} \hat{h}_{zz}} e^{-i \gamma_{j} \hat{h}_{x}}] \ket{0}^{\otimes N}$, where $\hat{h}_{zz} = \sum_{n = 1}^{N} \hat{\sigma}_{n}^{z} \hat{\sigma}_{n + 1}^{z}$ and $\hat{h}_{x} = \sum_{n = 1}^{N} \hat{\sigma}_{n}^{x}$ correspond to the first and second terms of Eq.~\eqref{eq:Hamiltonian} and the parameters $\beta_{i}$ and $\gamma_{i}$ are randomly chosen from a uniform distribution from $-\pi$ to $\pi$. The purpose of increasing the number of layers as the system size grows is to create enough randomness and entanglement of the random initial state. As shown in Fig.~\ref{ini_state}, the fidelity between the random initial state and ground state decays fast as the system size grows. However, as the system size increases, the well-prepared initial ones still have large overlaps with the ground state. This demonstration shows that prior knowledge of the Hamiltonian can help prepare a good quality initial state for a better algorithm performance.

\section{Continuous variable assisted strategy}
\label{Sec_Continuous_variable_assisted_algorithm}
In this section we introduce an alternative iterative quantum algorithm with the assistance of the continuous-variable qumode~\cite{zhang2020protocol}. In this strategy, the Gaussian function of the Hamiltonian is constructed by integration of unitary operators, which is completed by the entangling qubit and qumode since infinite integration naturally exists in the qumode state~\cite{lau2017quantum,zhang2021continuous,he2022inverse}. The qumode state can be presented as a Fock state $\sum_{n = 0}^{cut} c_{n} \ket{n}$ or a pair of conjugate quadrature states $\int_{-\infty}^{\infty} f_{p}(p) \ket{p} dp$ and $\int_{-\infty}^{\infty} f_{q}(q) \ket{q} dq$, such as momentum $\hat{p}$ and position $\hat{q}$ of photons. Consider a finite squeezed state $\ket{\phi} = s^{-1/2} \pi^{-1/4} \int_{-\infty}^{\infty} e^{-p^{2}/2 s^{2}} \ket{p} dp$, where $s$ is the squeezing factor; this qumode is finitely squeezed on momentum space and extended on position space.

The Gaussian filter operator can be constructed by first performing a unitary operator $e^{- i \hat{H} \hat{p}}$ on the hybrid qubit-qumode initial state $\ket{\psi_{i}} \ket{\phi}$ and then projecting the ancillary qumode on $\ket{\phi}$ by homodyne measurement~\cite{braunstein2005quantum}. The resulting state is shown as
\begin{equation}
    \begin{aligned}
        \braket{\phi| e^{-i \hat{H} \hat{p}} |\psi_{i}}\ket{\phi} 
        &= \frac{1}{s \sqrt{\pi}} \int_{-\infty}^{\infty} e^{-p^{2}/s^{2}} e^{-i p \hat{H}} \ket{\psi_{i}} dp \\
        &= \frac{1}{\sqrt{C}} \sum_{j} a_{j} e^{-s \lambda_{j}^{2}/2} \ket{\psi_{i}},
    \end{aligned}
\end{equation}
where $1/\sqrt{C} = (\sum_{j} a_{j}^{2} e^{- s \lambda_{j}^{2}})^{-1/2}$ is the normalization factor that determines the successful projection rate of the qumode. The unitary operator to entangle qubits and the qumode $e^{- i \hat{H} \hat{p}}$ can be decomposed into the evolution of local terms $e^{-i c_{l} \hat{h}_{t} \hat{p}/n}$ by Trotter decomposition~\cite{suzuki1991general}. Each element can be constructed by a single-qubit single-qumode gate and universal qubit gates, which are available on existing hybrid qubit-qumode platforms~\cite{gan2020hybrid, devoret2013superconducting}. The projection process is accomplished by squeezing the ancillary qumode and then measuring the vacuum state, which is a postselection process. The above equation indicates that both a greater squeezing factor and the scale of eigenvalues lead to higher accuracy but a lower successful projection rate. Moreover, the squeezing factor $s/2$ represents the inverse of the variance of the Gaussian filter $1/\sigma^{2}$ and the shift energy is encoded into the Hamiltonian.

Our strategy is to iteratively increase the shift energy applied with a fixed squeezing factor $s$ to estimate the ground state $\ket{\lambda_{0}}$ and its corresponding eigenvalue $\lambda_{0}$ until the desired accuracy is reached. Though the accuracy increases as the shift energy rises, the successful projection rate decreases and the algorithm is more sensitive to noises. It means the improvement of the results requires a cost of repetition, and this kind of improvement has an upper bound depending on the scale of noise in practice. As the successful projection probability decays, increasing the number of query samples leads to a greater probability of generating an erroneous state due to noise. For this reason, it is required for the users to set an appropriate maximum shift energy, whose scale depends on the hardware. Compared to the hybrid quantum-classical approach, this one does not need to measure many discrete terms but cannot adjust the parameters of the Gaussian filter classically. 

\begin{figure}
    \centering
        \begin{minipage}[b]{0.45\textwidth}
            \includegraphics[width=1\textwidth]{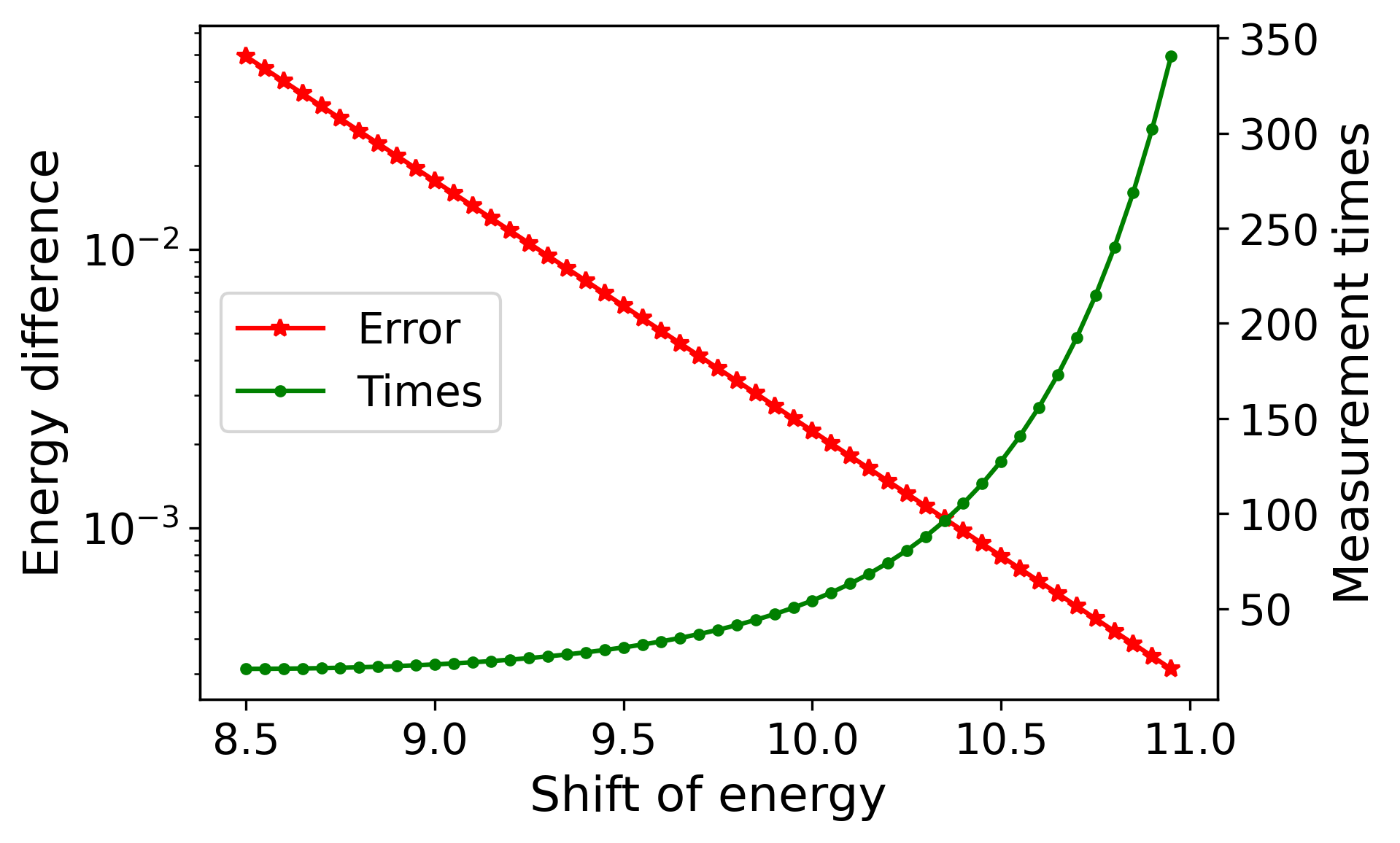}
        \end{minipage}
   
    \caption{Solution of an $N = 4$ qubit transverse-field Ising model by iterative continuous-variable-assisted strategy. The red line is the energy difference between the estimated energy and the exact ground-state energy, corresponding to the left axis. The green line is the required measurement times corresponding to the right axis.}
     \label{Ising_continuous_random_ini_E_vs_shift_E}
\end{figure}

We demonstrate this continuous-variable-assisted QGF algorithm by solving the approximate ground state and corresponding energy of a transverse-field Ising model with $J =1$, $g = 2$, and $N = 4$. The squeezing factor and maximum Fock state of qumode are set as $s = 1$ and a cutoff equal to $50$, and a random initial state is applied. As the exact ground state energy is $-8.543$ without applying external shift energy, it starts by shifting the exact ground-state energy to zero and then it is gradually increased. Fig.~\ref{Ising_continuous_random_ini_E_vs_shift_E} shows that the result error exponentially decays, but required measurement times exponentially grow as the shift energy rises. In practice, this procedure can stop when a result with the desired accuracy is obtained or the successful projection rate is too low.

Finallys, we briefly analyze the time complexity of this continuous-variable assisted quantum algorithm. This is similar to the hybrid quantum-classical one but replaces the $1/\sigma^{2}$ by $s$. Consider the case of solving a local gapped model under the same assumption as mentioned in Sec.~\ref{Sec_Time_complexity_analysis}. 
The resulting state after the Gaussian filter is performed is approximated as
$\ket{\psi_{f}} \approx \frac{1}{\sqrt{C}} (a_{0} e^{-s\lambda_{0}^{2}/2} \ket{\lambda_{0}} + \sqrt{1 - a_{0}^{2}} e^{-s(\lambda_{0} + \Delta)^{2}/2} \ket{\lambda_{\perp}})$.
The requirement of the ground-state energy scale is $\lambda_{0} = O(s^{-1} \log(\epsilon^{-1} a_{0}^{-2}))$ for solving an approximate ground-state energy with the desired accuracy $\epsilon$. The inverse square normalization factor $C = O(a_{0}^{2} e^{-s\lambda_{0}^{2}})$ presents the successful projection rate of the nonunitary operator. So the requirement of a query sample is $O(a_{0}^{-2} e^{s\lambda_{0}^{2}})$. Considering the Trotter decomposition, the circuit depth is $O(\epsilon^{-1} L^{3} a_{0}^{-2} e^{s\lambda_{0}^{2}})$ since the error results from Trotterization should be $O(\epsilon a_{0}^{2} e^{-s\lambda_{0}^{2}})$. The time complexity in total is $O(\epsilon^{-1} L^{3} a_{0}^{-4} e^{2s\lambda_{0}^{2}}) = O(\epsilon^{-1} L^{3} a_{0}^{-4} e^{2 s^{-1} \log^{2}(\epsilon^{-1} a_{0}^{-2})})$. It depends on the initial overlap and the squeezing factor.

\section{Conclusion}
\label{Sec_Conclusion_and_discussion}
We have proposed a hybrid quantum-classical algorithm for solving the approximate ground-state energy well suited to NISQ computers. By constructing a Gaussian filter as a linear combination of unitary operators and expressing the estimated energy as a weighted summation of state overlap, the best-performance coefficient can be obtained with the help of post-processing to exploit the quantum resources efficiently. This strategy solves an approximate ground-state energy with a time complexity that is a polynomial of the system size, desired accuracy, and inverse of overlap between the initial state and exact ground state. After solving the ground-state energy, any other expectation values of the observable for the ground-state can be estimated by replacing $\hat{H}$ with an observable operator $\hat{A}$ for exploring the ground-state properties. Moreover, the classical optimization strategy for solving the ground-state energy can also apply to other quantum filtering operators, such as the cosine filter (the example is shown in Appendix.~\ref{Sec_Comparison_to_the_cosine_filter_algorithm}).

We have shown the performance of our algorithm by solving the ground-state energy of a transverse-field Ising model with numeral simulations. In practice, this algorithm can be used iteratively by extending the evolution time series to obtain a higher accuracy result. Moreover, we have also simulated our algorithm in a noisy environment, which still performs well with the help of error mitigation without using any ancilla. We have given a comparison to the cosine filter both theoretically and numerically. The Gaussian filter and cosine filter have a similar performance under noise. A continuous-variable-assisted iterative quantum algorithm has also been presented, which constructs the Gaussian filter by qubit-qumode entanglement and solves the approximate ground state by iteratively increasing the applied shift energy. Its time complexity is strongly dependent on the initial overlap and squeezing factor. This alternative method is well suited for hybrid qubit-qumode quantum processors, and the numerical demonstration shows its good performance.

\textit{Note added:} Recently, we noticed three recent papers~\cite{keen2021quantum, wang2022state,apers2021quadratic} relevant to our work. Those works provide a complementary approach to the quantum Gaussian filter.

\section*{Acknowledgements}
This work was supported by the Key-Area Research and Development Program of Guangdong Province (Grant No. 2019B030330001), the National Natural Science Foundation of China (Grant No. 12005065), and the CRF grant of Hong Kong (Grant No. C6005-17G and No. C6009-20G).

\appendix
\section{Approximation error of Gaussian filter}
\label{Sec_Approximation_error_of_Gaussian_filter}
We briefly discuss the approximation error to construct quantum Gaussian filters. Considering a Fourier transformation of a Gaussian function with zero mean value, $f(h) = e^{-h^{2}/\sigma^{2}} = \frac{\sigma}{2\sqrt{\pi}} \int_{-\infty}^{\infty} e^{-\sigma^{2} y^{2}/4-ihy} dy$, a quantum Gaussian filter with zero shift-energy can be expressed as
\begin{equation}
    f(\hat{H}) = e^{-\hat{H}^{2}/\sigma^{2}} = \frac{\sigma}{2\sqrt{\pi}} \int_{-\infty}^{\infty} e^{-\sigma^{2} y^{2}/4} e^{-i \hat{H} y} dy.
\end{equation}
However, the Hamiltonian evolution for an infinitely long time is experimentally impossible. A cutoff of this integration $Y$ is necessary to implement the Gaussian filter practically. The function $f(h)$ is approximated by 
\begin{equation}
    \begin{aligned}
        f_{Y}(h) &= \frac{\sigma}{2\sqrt{\pi}} \int_{-Y}^{Y} e^{-\sigma^{2} y^{2}/4-ihy} dy \\
        &= \frac{1}{2} e^{-h^{2}/\sigma^{2}} [{\rm erf}(i \frac{h}{\sigma} + \frac{\sigma Y}{2}) + {\rm erf}(i \frac{h}{\sigma} - \frac{\sigma Y}{2})],
    \end{aligned}
\end{equation}
where ${\rm erf}(x)$ is the error function of $x$. By setting $a = \frac{h_{m}}{\sigma} = \frac{\sigma Y}{2} (Y = \frac{2 h_{m}}{\sigma^{2}})$ and substituting $h_{m}$ into $f_{Y}(h)$, the above function becomes
\begin{equation}
    f_{Y}(h_{m}) = \frac{1}{2} e^{-h_{m}^{2}/\sigma^{2}} [{\rm erf}(i a + a) + {\rm erf}(i a - a)].
\end{equation}
The Taylor series for it at $a \to \infty$ is
\begin{equation}
    e^{-h_{m}^{2}/\sigma^{2}} \{ 1 + \frac{1}{2\sqrt{\pi}a}[\sin(2a^{2}) - \cos(2a^{2})] + O(a^{-3}) \}.
\end{equation}
The difference $|f(h) - f_{Y}(h)|$ for $h = h_{m}$ is $O(\frac{\sigma}{h_{m}} e^{-h_{m}^{2}/\sigma^{2}})$. So, the approximate error $|f(h) - f_{Y}(h)|$ for $h \in [0, h_{m}]$ has an upper bound $O(\frac{\sigma}{h_{m}} e^{-h_{m}^{2}/\sigma^{2}})$, if a truncation of integration $Y = \frac{2h_{m}}{\sigma^{2}}$ is applied.

Next is the error resulting from discrete summation. We discretize the integration to a summation represented as
\begin{equation}
    f_{Y, \Delta y} = \frac{\sigma}{2\sqrt{\pi}} \sum_{j_{y} = -Y/\Delta y}^{Y/\Delta y} e^{-\sigma^{2} (j_{y} \Delta y)^{2}/4 - i h j_{y} \Delta y} \Delta y,
\end{equation}
where $Y/\Delta y$ is an integer. The difference between the integral and sum is
\begin{widetext}
\begin{equation}
    \label{Eq:discrete_appro}
    \begin{aligned}
        |f_{Y, \Delta y} - f_{Y}| &= \frac{\sigma}{2\sqrt{\pi}} \left |\int_{-Y}^{Y} e^{-\sigma^{2} y^{2}/4-ihy} dy -\sum_{j_{y} = -Y/\Delta y}^{Y/\Delta y} e^{-\sigma^{2} (j_{y} \Delta y)^{2}/4 - i h j_{y} \Delta y} \Delta y \right | \\
        &= \frac{\sigma}{2\sqrt{\pi}}  \left |\sum_{j_{y} = -Y/\Delta y}^{Y/\Delta y} (\int_{j_{y} \Delta y}^{(j_{y} + 1) \Delta y} e^{-\sigma^{2} y^{2}/4-ihy} dy - e^{-\sigma^{2} (j_{y} \Delta y)^{2}/4 - i h j_{y} \Delta y} \Delta y) \right |.
    \end{aligned}
\end{equation}
\end{widetext}
The first term expanded at $\Delta y = 0$ is
\begin{equation}
    e^{-\sigma^{2} (j_{y} \Delta y)^{2}/4 - ih j_{y} \Delta y} [ \Delta y - \frac{\Delta y^{2}}{4} (i 2 h + \sigma^{2} j_{y} \Delta y) + O(\Delta y^{3})] 
\end{equation}
Eq.\eqref{Eq:discrete_appro} becomes

\begin{equation}
    \begin{aligned}
        &|f_{Y, \Delta y} - f_{Y}| \\
        &\approx \frac{\sigma \Delta y}{8\sqrt{\pi}} \left | \sum_{j_{y} = -Y/\Delta y}^{Y/\Delta y} e^{-\frac{\sigma^{2} (j_{y} \Delta y)^{2}}{4} - ih j_{y} \Delta y} (2 i h + \sigma^{2} j_{y} \Delta y) \Delta y \right | \\
        &\approx \frac{\sigma \Delta y}{8\sqrt{\pi}} \left | \int_{-Y}^{Y} e^{-\sigma^{2} y^{2}/4 - ih y} (2 i h + \sigma^{2} y) dy\right | \\
        &\approx \frac{\sigma \Delta y}{2 \sqrt{\pi}} e^{-\sigma^{2} Y^{2}/4} \left | \sin(hY) \right |.
    \end{aligned}
\end{equation}
The discretization error is $O(\sigma \Delta y e^{-h_{m}^{2}/\sigma^{2}})$. So, the requirement of slice thickness is $\Delta y = O(\frac{1}{h_{m}})$ to constrain the error in $O(\frac{\sigma}{h_{m}} e^{-h_{m}^{2}/\sigma^{2}})$.

\begin{figure}
    \centering
    \subfigure[]{
    \begin{minipage}[b]{0.45\textwidth}
        \includegraphics[width=1\textwidth]{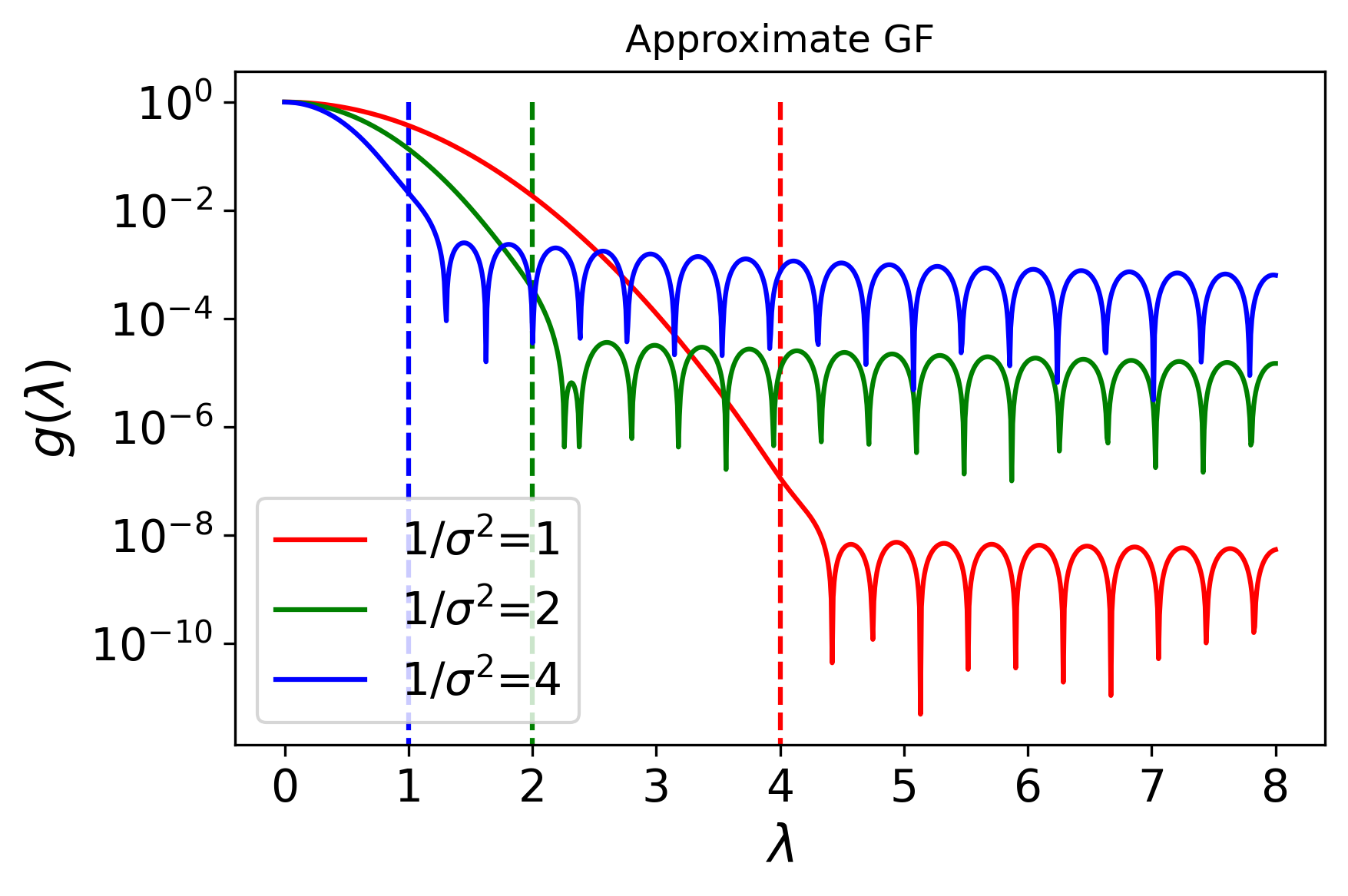}
        \label{fig_approx_GF_sig}
    \end{minipage}}
    \subfigure[]{
    \begin{minipage}[b]{0.45\textwidth}
        \includegraphics[width=1\textwidth]{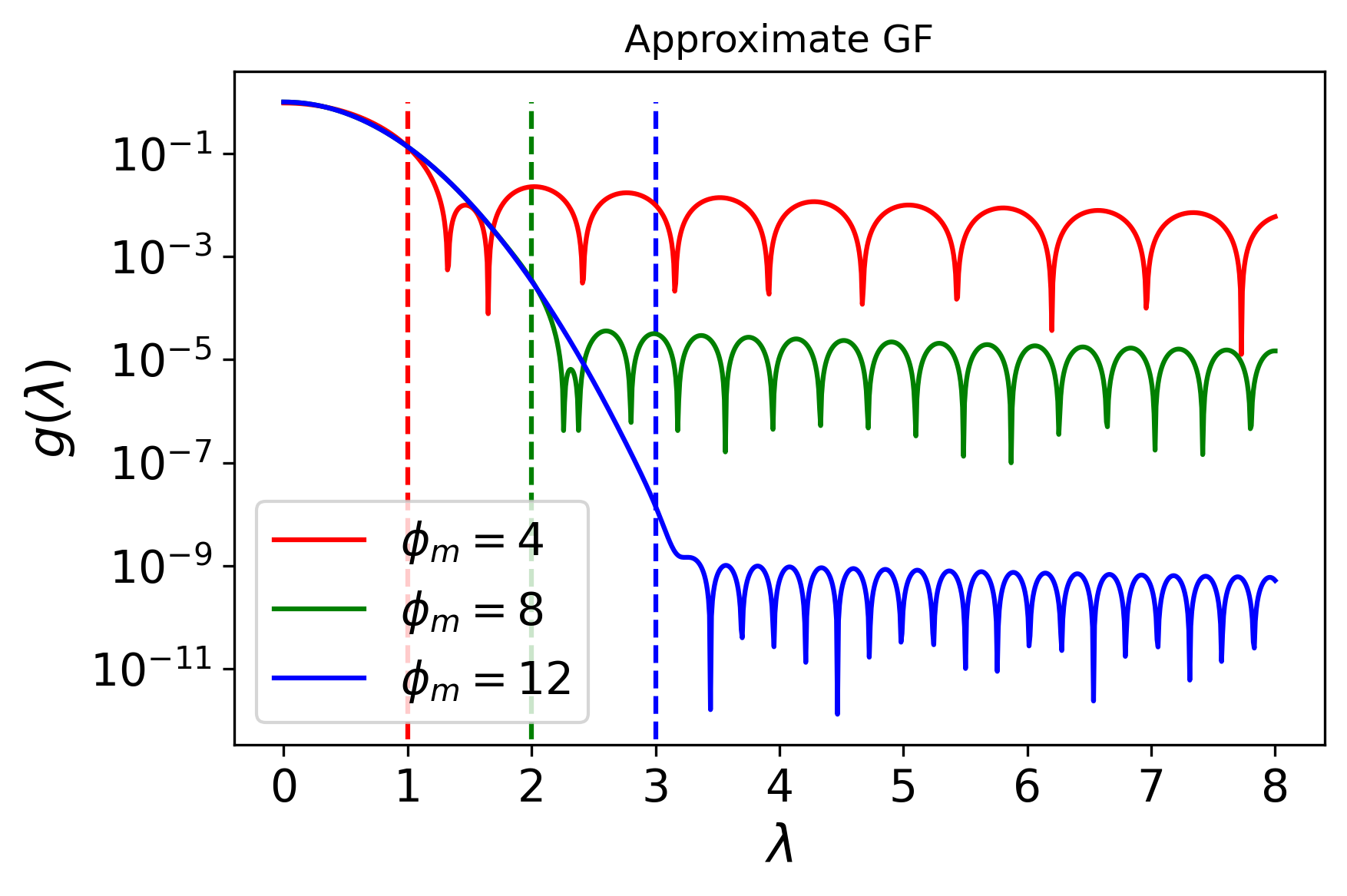}
        \label{fig_approx_GF_phase}
    \end{minipage}}
    
    \caption{Numerical demonstration of the approximate Gaussian functions for different settings. The slice thickness is set to a sufficient low scale $\Delta y = 0.16$. The vertical dashed lines correspond to $\lambda = \phi_{m} \sigma^{2}/2$. (a) Approximate Gaussian function for a fixed truncation $\phi_{m} = 8$ and three different inverse variances $1/\sigma^{2}$ = $1$, $2$, and $4$. (b) Approximate Gaussian function for a fixed inverse variance $1/\sigma^{2} = 2$ and three different truncations $\phi_{m}$= $4$, $8$, and $12$.}
    \label{fig_approx_GF}
\end{figure}

Finally, if we want to approximate a Gaussian function of the Hamiltonian $e^{-\hat{H}^{2}/\sigma^{2}}$ by a linear combination of unitaries for the eigenvalue $\lambda \in [0, \lambda_{m}]$ with an upper bound of error $O(\frac{\sigma}{\lambda_{m}} e^{-\lambda_{m}^{2}/\sigma^{2}})$, the maximum evolution time and slice thickness requirements are $\phi_{m} = 2 \lambda_{m}/\sigma^{2}$ and $O(1/\lambda_{m})$. Moreover, the addition shift energy of the QGF is completed by classically shifting the Gaussian function. So the well-approximated eigenvalue range for $e^{-(\lambda - \mu \hat{I})^{2}/\sigma^{2}}$ is $\lambda \in [\mu, \lambda_{m} + \mu]$ under the above conditions. Here we also numerically demonstrate the approximate Gaussian functions. We have set the discretization to a sufficiently low scale $\Delta y = 0.16$. We have plotted in Fig.~\ref{fig_approx_GF} the approximate Gaussian function for different parameters and $\lambda = \phi_{m} \sigma^{2}/2$ (vertical dashed lines). Fig.~\ref{fig_approx_GF_sig} shows the result for a fixed $\phi_{m} = 8$ and three values $1/\sigma^{2}$ = $1$, $2$, and $4$; and Fig.~\ref{fig_approx_GF_phase} shows the result for a fixed $1/\sigma^{2} = 2$ and three values $\phi_{m}$ = $4$, $8$, and $12$. Both show that the Gaussian filter well approximates the Gaussian function for $\lambda \in [0, \phi_{m} \sigma^{2}/2]$ and then oscillates.

\section{Requirement of Trotter number and statistical error}
\label{Sec_Requirement_of_Trotter_number_and_statistical_accuracy}
In the QGF algorithm, we filter out the unnecessary state components to converge the initial state into an approximate ground state. This filtering process leads to a decay problem in that the norm of the resulting state decreases while pursuing a higher accuracy. However, it makes the result sensitive to the overlap measurement error, including the error from Trotter decomposition and statistic error. Therefore, we will discuss in this appendix the decay problem and the requirement of the Trotter step and query sample for this problem.

We recall that our hybrid quantum-classical algorithm's formula for the estimated energy is represented as 
\begin{equation}
    \begin{aligned}
        \lambda_{\mu, \sigma}
        =&  \frac{\braket{\psi_{\mu, \sigma} | \hat{H} | \psi_{\mu, \sigma}}}{\braket{\psi_{\mu, \sigma} | \psi_{\mu, \sigma}}} \\
        =& \frac{\sum_{y, y^{\prime} = -M_{y}}^{M_{y}} b_{\mu, \sigma, y} b_{\mu, \sigma, y^{\prime}}^{\ast} \braket{\psi_{i}|\hat{H} e^{-i (t_{y} - t_{y^{\prime}}) \hat{H}}|\psi_{i}}}{\sum_{y, y^{\prime} = -M_{y}}^{M_{y}} b_{\mu, \sigma, y} b_{\mu, \sigma, y^{\prime}}^{\ast} \braket{\psi_{i}|e^{-i (t_{y} - t_{y^{\prime}}) \hat{H}}|\psi_{i}}},
    \end{aligned}
\end{equation}
where $b_{\mu, \sigma, y} = \frac{\sigma}{2 \sqrt{\pi}} e^{- (y \Delta_{y} \sigma)^{2}/4} e^{ i \mu (y \Delta_{y})}$. Both the denominator and numerator have $4M_{y} + 1$ unique terms, whose evolution time ranges from $-2\phi_{m}$ to $2\phi_{m}$, and the interval is $\Delta y$. Considering the approximate Gaussian filter mentioned in Sec.~\ref{Sec_Time_complexity_analysis} applied to the initial state, the normalization factor of the resulting state is $1/\sqrt{C} = [a_{0}^{2} e^{-2(\lambda_{m} - \Delta)^{2}/\sigma^{2}} + (1 - a_{0}^{2}) e^{-2\lambda_{m}^{2}/\sigma^{2}}]^{-1/2}$. The scale of the denominator is approximated to $a_{0}^{2} e^{-2(\lambda_{m} - \Delta)^{2}/\sigma^{2}}$ since $a_{0}^{2} e^{-2(\lambda_{m} - \Delta)^{2}/\sigma^{2}} \gg (1 - a_{0}^{2}) e^{-2\lambda_{m}^{2}/\sigma^{2}}$. The scale of the numerator is approximately $L$ times the denominator, which indicates the numerator term has a weaker condition.

If we pursue an accuracy of the result within $\epsilon$, the statistical error of each term should be $\epsilon a_{0}^{2} e^{-2(\lambda_{m} - \Delta)^{2}/\sigma^{2}}$. Since the weight has a zoom effect of $\frac{\sigma}{2\sqrt{\pi}} e^{- (y \Delta_{y} \sigma)^{2}/4}$ for the error of the $y$th term, the statistical error of the overlap in the $y$th term should be $\frac{2\sqrt{\pi}}{\sigma} \epsilon a_{0}^{2} e^{-2(\lambda_{m} - \Delta)^{2}/\sigma^{2}} e^{(y \Delta y)^{2} \sigma^{2}/4}$. So the number of total query samples is
\begin{equation}
    \begin{aligned}
        & \sum_{y = -2\phi_{m}/\Delta y}^{2\phi_{m}/\Delta y} \frac{\sigma}{2\sqrt{\pi}} \epsilon^{-1} a_{0}^{-2} e^{2(\lambda_{m} - \Delta)^{2}/\sigma^{2}} e^{-(y \Delta y)^{2} \sigma^{2}/4} \\
        &= \frac{\sigma}{2\sqrt{\pi}} \epsilon^{-1} a_{0}^{-2} e^{2(\lambda_{m} - \Delta)^{2}/\sigma^{2}} \Delta y^{-1} \sum_{y = -2\phi_{m}/\Delta y}^{2\phi_{m}/\Delta y} e^{-(y \Delta y)^{2} \sigma^{2}/4} \Delta y \\
        &\approx \frac{\sigma}{2\sqrt{\pi}} \epsilon^{-1} a_{0}^{-2} e^{2(\lambda_{m} - \Delta)^{2}/\sigma^{2}} \Delta y^{-1} \int_{y = -2\phi_{m}}^{2\phi_{m}} e^{-y^{2} \sigma^{2}/4} dy \\
        &= \epsilon^{-1} a_{0}^{-2} e^{2(\lambda_{m} - \Delta)^{2}/\sigma^{2}} \Delta y^{-1} {\rm erf}(\sigma \phi_{m}).
    \end{aligned}
\end{equation}
Considering the energy gap is independent of the system size $\Delta = O(1)$ and the slice fitness is $\Delta y = O(\lambda_{m}^{-1})$, the query complexity is $O(\epsilon^{-1} a_{0}^{-2} \lambda_{m} e^{2\lambda_{m}^{2}/\sigma^{2}})$ for obtaining a result with a fixed error $\epsilon$.

We consider the first-order Trotter decomposition~\cite{suzuki1991general} to implement the time evolution of the Hamiltonian on quantum computers. The Hamiltonian is a summation of local operators $\hat{H} = \sum_{l=1}^{L} c_{l} \hat{h}_{l}$; its time evolution is decomposed into the local operator rotations, i.e.,
\begin{equation}
    e^{-i\hat{H} t} \approx (\prod_{l=1}^{L} e^{-ic_{l}\hat{h}_{l}t/n})^{n}.
\end{equation}
The upper bound of gate counts constraining the error within $\epsilon$ is $O(L^{3} c_{max}^{2} t^{2}/\epsilon)$. As the maximum weight of the local operator $c_{max}$ is independent of the system size, it is neglected. The $y$th term has an evolution time $t = y \Delta y$, and the amplitude of its corresponding weight is $\frac{\sigma}{2\sqrt{\pi}} e^{-(y \Delta y)^{2} \sigma^{2}/4}$. The error resulting from Trotter decomposition for $e^{-i\hat{H}t}$ should be $O(\epsilon \sigma^{-1} a_{0}^{2} e^{-2\lambda_{m}^{2}/\sigma^{2}} e^{t^{2}\sigma^{2}/4})$. Therefore, the gate counts to $O(\epsilon^{-1} L^{3} t^{2} \sigma a_{0}^{-2} e^{2\lambda_{m}^{2}/\sigma^{2}} e^{-t^{2} \sigma^{2}/4})$ for $e^{-i\hat{H}t}$, where the absolute value of the evolution time ranges from $0$ to $2\phi_{m} = 4\lambda_{m}/\sigma^{2}$. If $\sigma \leq 2\lambda_{m}$, the maximum value of the gate counting is $O(\epsilon^{-1} L^{3} \sigma^{-1} a_{0}^{-2} e^{2\lambda_{m}^{2}/\sigma^{2}})$ when $t = 2/\sigma$; if $\sigma > 2\lambda_{m}$, the maximum value of the gate counting is $O(\epsilon^{-1} L^{3} \lambda^{2} \sigma^{-3} a_{0}^{-2} e^{\lambda_{m}/\sigma^{2}})$ when $t = 4\lambda_{m}/\sigma^{2}$.

\section{Comparison to the cosine filter algorithm}
\label{Sec_Comparison_to_the_cosine_filter_algorithm}
In this section, we compare the cosine filter algorithm~\cite{lu2021algorithms,ge2019faster} to our Gaussian filter for solving the ground-state energy theoretically and numerically. In their work, a cosine filter is constructed by the LCU lemma to solve the energy spectrum of the Hamiltonian. The cosine-filter operator is approximated to a Gaussian function of the Hamiltonian. We will first give a brief analysis of the required quantum resource for constructing an approximate Gaussian function of the Hamiltonian by their formula. Next we will apply our classical optimization method to the cosine filter to compare their performances for solving the ground-state problem.

\begin{figure}
    \centering
    \subfigure[]{
    \begin{minipage}[b]{0.45\textwidth}
        \includegraphics[width=1\textwidth]{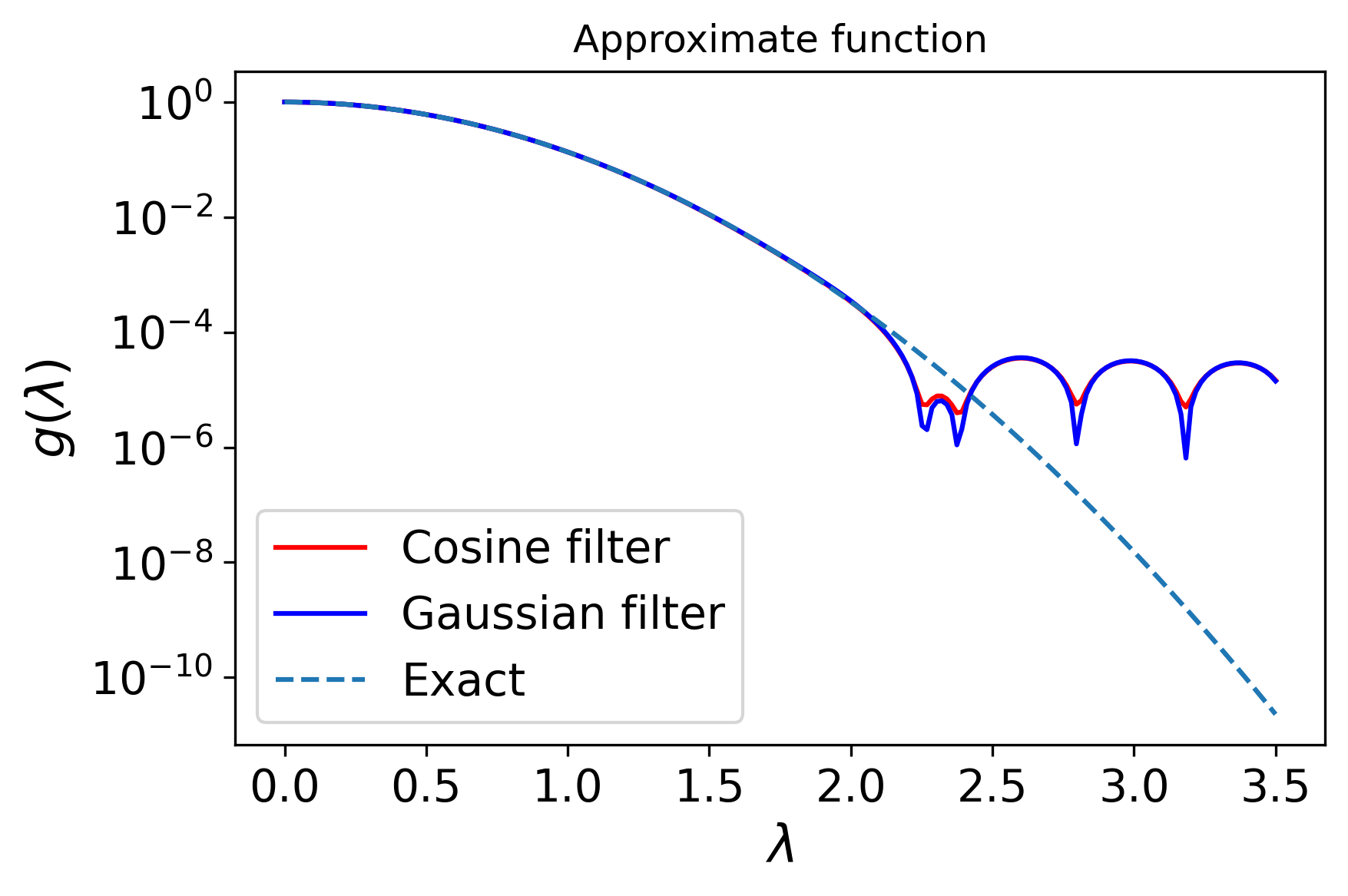}
        \label{fig_approx_GF_CF}
    \end{minipage}}
    \subfigure[]{
    \begin{minipage}[b]{0.45\textwidth}
        \includegraphics[width=1\textwidth]{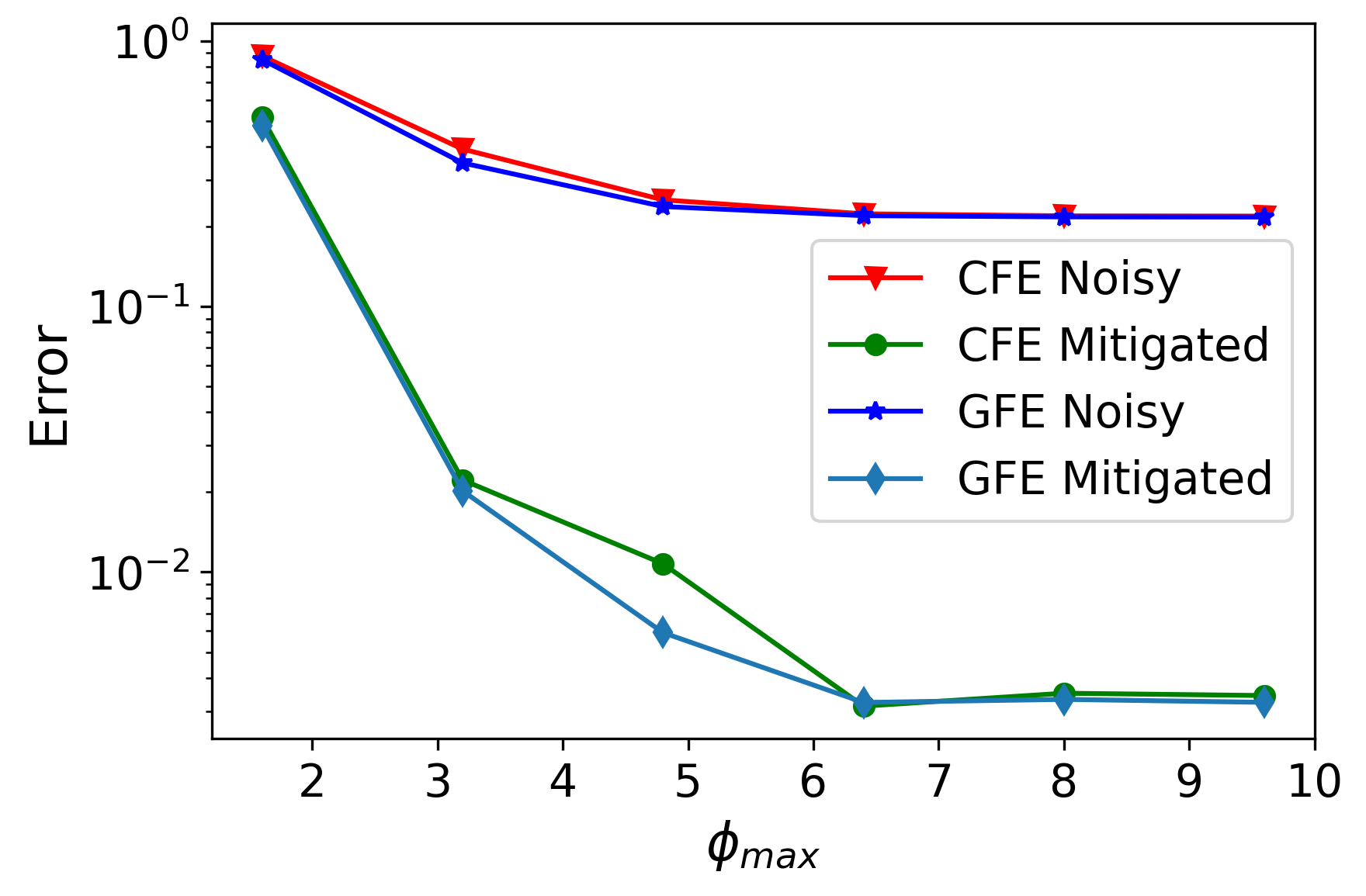}
        \label{fig_Ising_bit_error_compariso}
    \end{minipage}}
    
    \caption{Numerical comparison between the cosine filter and Gaussian filter. (a) Gaussian function of the eigenvalue approximated by the Gaussian filter and cosine filter, and the exact Gaussian function. We set the variance of the Gaussian $\sigma^{2} = \frac{1}{2}$, and the maximum evolution $\phi_{m} = 8$. The blue solid line, red solid line, and light blue dashed line correspond to the Gaussian filter, cosine filter, and exact Gaussian function, respectively. Both of the filters can approximate well the Gaussian function $e^{-2\lambda^{2}}$ for $\lambda \in [0, 2]$ with a maximum evolution time $\phi_{m} = 2h_{m}/\sigma^{2} = 8$. (b) Numerical result for solving the ground state of a four-qubit transverse-field Ising model for bit noise with $p_{b} = 0.0001$ after each gate. The red and blue lines are the results solved by the cosine filter and Gaussian filter in the noisy environment, respectively. The green and light blue lines are the results solved by the cosine filter and Gaussian filter with an error mitigation strategy, respectively.}
    \label{fig_comparision}
\end{figure}

The cosine-filtering operator is defined as a cosine function of the Hamiltonian and it is approximated to a Gaussian function of the Hamiltonian for sufficiently large power, i.e.,
\begin{equation}
    \begin{aligned}
        P_{\delta}(E) = \left [\cos(\frac{\hat{H} - E}{L}) \right ]^{L^{2}/\delta^{2}} \approx e^{-(\hat{H} - E)^{2}/2\delta^{2}},
    \end{aligned}
\end{equation}
where $L$ is the number of local operators on $\hat{H}$, and $L^{2}/\delta^{2}$ is set as an even number. In the original algorithm, $L$ is used to constrain the eigenvalues so that $\lVert(\hat{H} - E)/L\rVert_{\infty} \leq 1$, $E$ plays the role of scanning the energy spectrum, and $\delta$ corresponds to the width of the Gaussian function. The cosine-filtering operator is constructed by a linear combination of unitaries
\begin{equation}
    P_{\delta}(E) \approx \sum_{y = - x L/2\delta}^{x L/2\delta} c_{y} e^{i2y E/L} e^{-i2y \hat{H}/L},
\end{equation}
where $c_{y} = 2^{-L^{2}/\delta^{2}} \begin{pmatrix} L^{2}/\sigma^{2} \\ L^{2}/2\sigma^{2} - y
\end{pmatrix}$ and $x$ determines the approximation error. The parameters of the cosine-filtering operators $\delta$, $L$, and $x$ are set beforehand depending on the system and required accuracy. Here $x$ constrains the approximation error to $2 e^{-x^{2}/2}$ for $\lVert(\hat{H} - E)/L \rVert_{\infty} \leq 1$. It gives a clear requirement of quantum resources; the maximum evolution time is $\phi_{m} = x/\delta$ and the time interval is $\Delta t = 2/L$.

We numerically simulate the approximate Gaussian function of eigenvalues.
If we set the parameters $L = \lambda_{m}$, $\delta = \sigma/\sqrt{2} $, $E = 0$ and $x = \sqrt{2} \lambda_{m}/ \sigma$, the cosine-filtering operator can well approximate the Gaussian function $e^{-\hat{H}^{2}/\sigma^{2}}$ for eigenvalues $h \in [0, \lambda_{m}]$ with an accuracy $O(2e^{-\lambda_{m}^{2}/\sigma^{2}})$. The corresponding maximum evolution time is $O(2h_{m}/\sigma^{2})$, and the time interval is $\Delta t = O(\lambda_{m}^{-1})$. Recalling the requirement of quantum resource to construct the Gaussian filter mentioned on Appendix~\ref{Sec_Approximation_error_of_Gaussian_filter}, it approximates $e^{-h^{2}/\sigma^{2}}$ for $h \in [0, \lambda_{m}]$ with an accuracy $O(\frac{\sigma}{\lambda_{m}} e^{-\lambda_{m}^{2}/\sigma^{2}})$, where the maximum evolution is $O(2\lambda_{m}/\sigma^{2})$, and the slice fitness is $\Delta t = O(\lambda_{m}^{-1})$. They have similar query and circuit complexities. We also demonstrate the function of eigenvalues approximated by the cosine filter and Gaussian filter for the same time interval and maximum evolution time. As shown in Fig.~\ref{fig_approx_GF_CF}, both the Gaussian filter and the cosine filter can well approximate $e^{-2\lambda}$ for $\lambda \in [0, 2]$ with a maximum evolution $\phi_{m} = 2\lambda_{m}/\sigma^{2} = 8$.

We apply the classical optimization method proposed herein to the cosine-filtering operator that optimizes the parameters $\delta$ and $E$ to minimize the estimated energy under fixed discrete parameters $\Delta y$ and $\phi_{m}$. We give a numerical demonstration of solving the ground-state energy of a four-qubit transverse field Ising model mentioned in Sec.~\ref{Sec_Demonstration} in a quantum bit-flip environment with a probability $p_{b} = 0.0001$ after each quantum gate. We set the Trotter number to $20$ for a single discrete evolution time $\Delta y = 0.16$, and the maximum evolution time $\phi_{m}$ from $1.6$ to $9.6$. The classical scanning parameters ranges are $\mu \in [0, -0.5]$ and $1/\sigma^{2} \in [1, 3]$. As shown in Fig.~\ref{fig_Ising_bit_error_compariso}, the cosine filter and Gaussian filter give similar results. In summary, the cosine filter and the Gaussian filter have similar performances in both approximating the Gaussian function and solving ground-state energy. This indicates that our postprocess optimization suits other quantum filtering operators well.

\bibliography{reference.bib}

\end{document}